\documentclass{aa}  

\usepackage{graphicx}
\usepackage{stfloats}
\usepackage{txfonts}

\usepackage[colorlinks=true,allcolors=blue]{hyperref}

\usepackage{threeparttable}
\usepackage{makecell} 
\usepackage[dvipsnames]{xcolor}
\begin{document}

   \title{Spectroscopic characterisation of gravitationally lensed stars at high redshifts}

   \author{Emma Lundqvist\inst{\ref{UU-O}}
          \and
          Erik Zackrisson\inst{\ref{UU-O},\ref{SCAS}}
          \thanks{Corresponding author: erik.zackrisson@physics.uu.se}
		\and
			Calum Hawcroft\inst{\ref{stsci}}
		\and 
		      Anish~M.~Amarsi\inst{\ref{UU-T}}
        \and
                Brian Welch\inst{\ref{Goddard},\ref{Maryland}}
          }

   \institute{\label{UU-O}Observational Astrophysics, Department of Physics and Astronomy, Uppsala University, Box 516, SE-751 20 Uppsala, Sweden
    \and
     \label{SCAS}Swedish Collegium for Advanced Study, Linneanum, Thunbergsv\"a{}gen 2, SE-752 38 Uppsala, Sweden
     \and
    \label{stsci}Space Telescope Science Institute, 3700 San Martin Drive, Baltimore, MD 21218, USA
    \and
    \label{UU-T}Theoretical Astrophysics, Department of Physics and Astronomy, Uppsala University, Box 516, SE-751 20 Uppsala, Sweden
    \and
    \label{Goddard}Observational Cosmology Lab, NASA Goddard Space Flight Center, Greenbelt, MD 20771, USA
    \and
    \label{Maryland}Department of Astronomy, University of Maryland, College Park, 20742, USA}

   \date{Received ...; accepted ...}

  \abstract{Deep imaging of galaxy cluster fields have in recent years revealed tens of candidates for gravitationally lensed stars at redshifts $z\approx 1$--6, and future searches are expected to reveal highly magnified stars from even earlier epochs. Multi-band photometric observations may be used to constrain the redshift, effective temperature $T_\mathrm{eff}$, and dust attenuation along the line of sight to such objects. When combined with an estimate of the likely magnification, these quantities may be converted into a constraint on the stellar luminosity and, for an adopted set of stellar evolutionary tracks, the initial stellar mass. Further characterization is, however, difficult without spectroscopic observations, which at the typical brightness levels of high-redshift lensed stars becomes extremely challenging for even the largest existing telescopes. Here, we explore what spectral features one can realistically hope to detect in lensed stars with peak brightness in the range 26--28 AB mag, $T_\mathrm{eff}= 4000$--50\,000 K, and redshifts $z=$1--10, using spectroscopy with the \textit{James Webb} Space Telescope (JWST) and the forthcoming Extremely Large Telescope (ELT). We find that a majority of detectable lines appear in the rest UV-range for stars with $T_\mathrm{eff}\geq15\,000$ K. The strongest detectable spectral lines are the \ion{C}{IV} $\lambda 1550$$\AA$ line and the \ion{Si}{IV} $\lambda\lambda$1393, 1403$\AA$-doublet at $T_\mathrm{eff}=30\,000$ K. For lower temperatures, the calcium H- and K-lines at $T_\mathrm{eff}=6000$ K are among the most readily detectable. In limited wavelength ranges, ELT is expected to provide more sensitive spectroscopic observations, and with higher resolution than JWST. We find that variations of both mass loss rate and metallicity lead to noticeable effects in the detectability of certain spectral lines with both JWST and ELT.}


   \maketitle
%

\section{Introduction}
In recent years, tens of candidates for gravitationally lensed stars at cosmological distances (redshifts $z\approx 1$--6) have been detected in galaxy cluster fields \citep[][]{Kelly18,Chen19,Kaurov19,Welch22a,Chen22,Kelly22,Diego23a,Meena23,Yan23,Diego23b,Fudamoto24}.  
Studies of gravitationally lensed stars hold the potential to constrain the properties and evolution of high-mass stars at low metallicities and at high redshifts, in addition to probing the stellar initial mass function at earlier epochs in the history of the Universe. Observations of lensed stars can furthermore constrain the properties of dark matter in the galaxy cluster responsible for the combination of macro- and microlensing that renders the star detectable \citep{Dai18,Oguri18,Dai20,Williams23,Diego23b}. The so-far elusive, chemically pristine ``Population III'' stars may under certain conditions also turn up in samples of lensed stars \citep{Windhorst18,Zackrisson23}, but it currently remains unclear how these would be distinguished from their metal-enriched counterparts.

Characterising lensed stars is challenging due to the faint magnitudes (typically $\gtrsim 26$ AB magnitudes) at which they are detected \citep[but see][for a potential exception]{Diego22}. By fitting the observed photometric spectral energy distribution (SED) of the lensed star to suitable models \citep[e.g.][]{Welch22b,Meena23,Diego23b}, it is possible to set constraints on the redshift\footnote{Although the best redshift measurements usually come from photometric and spectroscopic observations of the gravitational arc in which the lensed star appears, i.e. its host galaxy)}, the effective temperature ($T_\mathrm{eff}$) of the star and occasionally the amount of dust attenuation along the line of sight. Admittedly, there is a degeneracy between $T_\mathrm{eff}$ and the dust attenuation \citep[see][for an example]{Furtak24}, and $T_\mathrm{eff}$ is generally difficult to determine from photometry alone, especially at high temperatures \citep{Zackrisson23}. By combining these quantities with an estimate on the gravitational magnification $\mu$ (often in the range $\mu\sim 10^2$--$10^4$), the bolometric luminosity of the lensed star may be estimated and, via an assumed stellar evolutionary track, its initial mass. However, in some of the lensed stars studied to date, it seems that the observed SEDs may feature light from at least two stars with different $T_\mathrm{eff}$ \citep{Welch22b,Diego23b,Furtak24}, which severely complicates the analysis and leads to degenerate solutions in the fitting procedure.

To measure additional stellar parameters, most pressing the metallicity or individual chemical abundances, would require additional constraints from spectroscopy. This is, however, extremely challenging at the typical brightness levels of high-redshift lensed stars ($\gtrsim 26$ AB mag), even with the largest telescopes currently available, since the signal-to-noise ratio (S/N) in the spectral continuum tends to be much too low to allow weak absorption lines to be detected. So far, absorption features have only been detected from one lensed star candidate -- the anomalously bright $z\approx 2$ object Godzilla at $m\approx 22$ AB mag studied with MUSE \citep{Diego22}. No similar features at high S/N have been detected in the \textit{James Webb} Space Telescope (JWST) spectra of the other two stars for which spectroscopic observations have been performed (\citealt{Furtak24}, Welch et al., in prep.) -- the \citet{Meena23} $z\approx 5$ object at $m\approx 27$--28 AB mag and the $z\approx 6$ object Earendel \citep{Welch22a} at $m\approx 27$ AB mag. However, there are tentative signs of \ion{Ca}{II} H and K absorption seen at low S/N in the \citet{Furtak24} spectrum. For lensed stars at $\gtrsim 26$ AB mag, many of the standard spectral diagnostics used in studies of nearby stars are much too faint to be detectable within reasonable exposure times, and characterization of such objects will therefore need to rely on a small number of the very strongest absorption features.

Here, we investigate what spectral features one may hope to detect at different $T_\mathrm{eff}$ in the spectra of high-mass stars magnified to $\approx 26$--28 AB magnitudes with either the JWST or the Extremely Large Telescope (ELT). We present our procedure for generating JWST and ELT mock spectra of lensed, high-redshift stars in Sect.~\ref{sec:simulations}. Sect.~\ref{sec:features} outlines what spectral features are deemed the most likely to be detected at different $T_\mathrm{eff}$ and $m_\mathrm{AB}$. Their potential use in assessing mass loss rates, abundances, and metallicities of lensed stars are discussed in Sect.~\ref{sec:discussion}, along with a number of uncertainties concerning the astrophysics underlying these features. Our findings are summarised in Sect.~\ref{sec:conclusions}.


\section{Simulating observed spectra of high-redshift lensed stars}
\label{sec:simulations}

Throughout this paper, we consider stellar mock spectra for eight effective temperatures ($T_\mathrm{eff}=4000$K, 6000K, 8000K, 10\,000K, 15\,000K, 20\,000K, 30\,000K, 50\,000K) at four redshifts ($z=1, 3, 6, 10$) as representative of most lensed stars. These effective temperatures were chosen since most of them are, according to \cite{Zackrisson23}, likely to appear in a sample of lensed stars. Observations also indicate that this range of effective temperatures is reasonable.

\subsection{Stellar atmosphere model grids}
We simulate observed spectra based on stellar atmosphere models from two different spectral templates. For low temperature stars, we use the BOSZ stellar atmosphere models \citep{Bohlin17}, based on the ATLAS-APOGEE ATLAS9 code \citep{Meszaros12}. BOSZ offers models for a large range of effective temperatures, resolving powers, as well as metallicities. Here we restrict our selection to $T_\mathrm{eff}=4000-10\,000$ K, $R=50\,000$, and use $[\mathrm{M/H}]=-1$ (with solar abundances) unless otherwise indicated. The $\alpha$-abundance is kept equal to that of the Sun, where the $\alpha$-elements are defined as O, Ne, Mg, Si, S, Ca, and Ti. The BOSZ models include no stellar winds, and also no chromospheres. One parameter that is able to affect the width of the lines is microturbulence, here with a velocity of $2.0 \mathrm{\,\,km\,s^{-1}}$, and not possible to vary for this grid. 

The projected rotational velocity and macroturbulence are also constant for the grid, both being $0 \mathrm{\,\,km\,s^{-1}}$. These parameters do not strongly influence the present study, as it is based on equivalent widths, on which the stellar rotation and macroturbulence have no direct impact to first order. However, extreme values could reduce the significance of detections, since very broadened lines and noise will become indistinguishable. In any case, the low resolving power of the instruments are expected to dominate the overall macroscopic broadening for realistic values of rotational velocity and macroturbulence. 

For higher temperature stars ($T_\mathrm{eff}=15\,000-50\,000$ K), we instead use OB-star models generated by the Potsdam Wolf-Rayet (PoWR) code \citep{Hainich19}. PoWR includes stellar winds in the modelling, and provides model grids for different mass loss rates. Unless otherwise specified, a moderate mass loss rate ($\log\dot{M}\left[\mathrm{M_\odot yr^{-1}}\right]\sim-6.7$) is applied to the stellar atmosphere models used here. Additionally, PoWR offers three choices for the metallicity of the OB-stars, and we adopt a metal content equal to that of the Small Magellanic Cloud (SMC), with $[\mathrm{M/H}]\sim-0.845$. The downloadable PoWR-files of stellar spectra have a resolving power of $R=160\,000$, and a microturbulence of $10  \mathrm{\,\,km\,s^{-1}}$ is used. 

Both model grids also allow for choosing the surface gravity, $\log(g)$. Here the lowest possible value of $\log(g)$ is always selected, since supergiants and hypergiants are more likely to be detectable through lensing than main-sequence stars \citep{Zackrisson23}. Empirically, current samples of lensed, high-mass stars also seem to be heavily weighted in favour of post-main sequence temperatures. 

Choosing representative stellar models for observable lensed stars is difficult, as the selection function is complicated \citep{Zackrisson23}. Although the chosen stellar atmosphere models hopefully represent lensed stars well, they are still not able to encompass every possible type of lensed star. More extreme values of $\dot M$, abundances, and $\log(g)$ could possibly alter the results.  We discuss the possible consequences of more extreme values of these parameters in Sects.~\ref{winds}, \ref{abundance}, and \ref{logg} respectively.

\subsection{JWST modelling}
We generate mock spectroscopic JWST observations based on the BOSZ and PoWR stellar models using the JWST exposure time calculator (ETC) v.3.0 \citep[see e.g.][]{Pontoppidan16}. With the Near Infrared Spectrograph (NIRSpec) grating/filter pairs G140M/F100LP, G235M/F170LP, and G395M/F290LP, we attain a resolution of $R\sim1000$, automatically adjusted with flux-conserving resampling by choosing a grating in the ETC, and cover the observed wavelength range $\lambda\sim0.97$--5.1 $\mu$m once the three grating/filter spectra are merged. The high resolution NIRSpec gratings, with $R\sim2700$, were not considered, as the over-all sensitivity decreases with resolution. Here, we adopt an extremely long exposure time of 50 hours per grating/filter and assume that the observations are carried out using the fixed slit (FS) S200A1 ($0.2 \times 3.2$ arcsec). The fixed slit is chosen since this produces the highest possible S/N for a point-like source, thereby setting a hard limit on what spectral features can realistically be detected\footnote{We note, however, that the NIRSpec MSA option is more practical for observations of cluster lensed fields, since many targets can then be observed simultaneously, and would result in S/Ns that are only marginally worse.}. Default coordinates and a medium background were used for the simulated observations. In the ETC, the spectra are normalised to their peak Near Infrared Camera (NIRCam) broadband filter flux (in the 26--28 AB mag range), which we hereafter refer to as the peak magnitude of the spectra. Based both on theoretical considerations and current samples of lensed stars, peak magnitudes $<26$ AB mag are expected to be exceedingly rare (and, if caused by microlensing, also of very short duration) and peak magnitudes $>28$ mag would result in incredibly noisy spectra, even with very long exposure times on the ELT. Mock spectra are generated at various redshifts, to investigate how the redshift affects the detectable spectral features of the lensed stars. 

Apart from determining the rest-frame wavelength range covered by NIRSpec, the redshift is important for its impact on the spectral resolution at which the rest-frame spectrum is captured by the instrument. Since the spectral resolution of NIRSpec increases with wavelength within each $R\sim 1000$ grating, and across the full wavelength range ($\lambda\sim0.97$--5.1 $\mu$m) of NIRSpec when spectra from several gratings are combined, a specific spectral feature seen at a redshift that places it towards the upper end of the NIRSpec wavelength interval can formally be measured at a higher spectral resolution (by a factor of $\lesssim 2$) than when seen at a redshift that places it at the lower end. However, since the sensitivity also changes in a non-monotonic way across the gratings, this does not necessarily make higher-redshift features more easily detectable than low-redshift ones, even at fixed AB magnitude of the source.

When plotting the JWST/NIRSpec mock spectra, the results from all three grating/filter pairs are merged together for the purpose of demonstrating where in the total NIRSpec wavelength range interesting features are likely to be detectable. Obtaining an observed spectrum that covers the full plotted range would hence require a total exposure time of $3\times 50$ hours, but since the number of detectable features is typically small, one would in more realistic observing campaigns likely focus on the single grating/filter that covers the features of interest, based on the photometric redshift of the target.

\subsection{ELT modelling}
Mock spectra of lensed stars observed with the Extremely Large Telescope (ELT) are also generated as part of this study, using the specifications of the High Angular Resolution Monolithic Optical and Near-infrared Integral field spectrograph (HARMONI). Two near-infrared bands are available for HARMONI, I+z+J covering $\lambda\sim 0.811$--1.369 $\mu$m and the H+K band covering $\lambda\sim1.450$--2.450 $\mu$m. HARMONI sensitivity estimates are based on two different options for adaptive optics (AO), where we here adopt the laser tomography AO (LTAO) option, since this does not require a bright natural guide star close to the target field and is therefore the most relevant setup for observations of lensed stars in cluster fields. Spectroscopy with HARMONI achieves a S/N of 5 in the continuum with an exposure time of $t_\mathrm{exp}=5$ hours, $R\sim3300$, a spatial sampling of 20 mas and LTAO at the calibration magnitude $m_\mathrm{AB}\approx 27.3$ \citep{Thatte16}\footnote{See also \url{https://harmoni-elt.physics.ox.ac.uk/index.html}}. Based on these specifications, we derive our S/N estimates per spectral bin on the assumption that S/N scales as $\mathrm{S/N}\propto F_\nu$, and that the detection level $F_{\nu,\mathrm{detection}}$ at fixed S/N scales as $F_{\nu,\mathrm{detection}} \propto 1/\sqrt{t_\mathrm{exp}}$ (in rough agreement with results from the ELT Spectroscopic Exposure Time Calculator v.6.4.0 at AB mag $\gtrsim 26$ in the near-IR).

\begin{table*}[!htbp] 
\centering
    \begin{threeparttable}
    \caption{Detectable metal lines in JWST/NIRSpec mock spectra at 26 AB mag for different effective temperatures and redshifts.}
    \label{tab:MetalLines}
        \begin{tabular}{c|c|llll}
        \hline \hline 
        \noalign{\smallskip}
                         $T_\mathrm{eff}$ [K]    & $\log(g)$ [cgs]   & $z=1$ & $z=3$ & $z=6$                                                                                                                                                  & $z=10$  \\ \noalign{\smallskip} 
                                \Xhline{0.8pt}
        4000  & 0.0 & No detectable lines     &   No detectable lines    &    No detectable lines &  No detectable lines      \\ \hline
        6000  & 0.0 & No detectable lines     &   \begin{tabular}[c]{@{}l@{}}\ion{Ca}{II} $\lambda$3934$\AA$\\ \ion{Ca}{II} $\lambda$3969$\AA$\end{tabular}    &    \begin{tabular}[c]{@{}l@{}}\ion{Ca}{II} $\lambda$3934$\AA$$^{a}$\\ \ion{Ca}{II} $\lambda$3969$\AA$$^{a}$\end{tabular}                                                                                                                                                     &  \begin{tabular}[c]{@{}l@{}}\ion{Ca}{II} $\lambda$3934$\AA$$^{a}$\\ \ion{Ca}{II} $\lambda$3969$\AA$$^{a}$\end{tabular}      \\ \hline
        
        8000  & 1.0 &  No detectable lines   &  No detectable lines     &  No detectable lines                                                                                                                                                      &    No detectable lines     \\ \hline
        10\,000 & 2.0 & No detectable lines    &  No detectable lines     &  No detectable lines                                                                                                                                                &   No detectable lines     \\ \hline
        15\,000 & 2.0 & No detectable lines     &   No detectable lines    &   \begin{tabular}[c]{@{}l@{}}\ion{Si}{IV} $\lambda$1393$\AA$  \\ \ion{C}{IV}  $\lambda$1550$\AA$$^{b}$\\ \ion{N}{IV} $\lambda$1719$\AA$$^{b}$\\ \ion{Al}{III} $\lambda$1854$\AA$ \\ \ion{Mg}{IV} $\lambda$1894$\AA$$^{b}$\end{tabular}                                                                                                                                                     &   \begin{tabular}[c]{@{}l@{}}\ion{Si}{II} $\lambda$1265$\AA$\\ \ion{Si}{III} $\lambda$1295$\AA$\\ \ion{C}{II}  $\lambda$1335$\AA$\\ \ion{Si}{IV} $\lambda$1393$\AA$\\ \ion{Si}{IV} $\lambda$1403$\AA$\\ \ion{C}{IV} $\lambda$1550$\AA$$^{b}$\\ \ion{N}{IV} $\lambda$1719$\AA$$^{b}$\\ \ion{Al}{III} $\lambda$1854$\AA$ \\\ion{Mg}{IV} $\lambda$1894$\AA$$^{b}$\end{tabular}     \\ \hline
        
        20\,000 & 2.4 & No detectable lines      & No detectable lines       & \begin{tabular}[c]{@{}l@{}}\ion{Si}{IV} $\lambda$1393$\AA$\\ \ion{Si}{IV} $\lambda$1403$\AA$\\ \ion{C}{IV} $\lambda$1550$\AA$$^{b}$\\ \ion{N}{IV} $\lambda$1719$\AA$$^{b}$\\ \ion{Al}{III} $\lambda$1854$\AA$\\ \ion{Mg}{IV} $\lambda$1894$\AA$$^{b}$ \end{tabular} &   \begin{tabular}[c]{@{}l@{}}\ion{Si}{III} $\lambda$1295$\AA$\\ \ion{C}{II}  $\lambda$1335$\AA$\\ \ion{Si}{IV} $\lambda$1393$\AA$\\  \ion{Si}{IV} $\lambda$1403$\AA$\\ \ion{C}{IV} $\lambda$1550$\AA$$^{b}$\\ \ion{Al}{III} $\lambda$1854$\AA$ \\\ion{Mg}{IV} $\lambda$1894$\AA$$^{b}$\end{tabular}      \\ \hline
        
        30\,000  & 3.2 &No detectable lines      & No detectable lines      & \begin{tabular}[c]{@{}l@{}}\ion{Si}{IV} $\lambda$1403$\AA$\\ \ion{C}{IV} $\lambda$1550$\AA$$^{a, b}$\\ \ion{Fe}{IV} $\lambda$1631$\AA$$^{b}$\\ \ion{N}{IV} $\lambda$1719$\AA$$^{b}$\end{tabular}                                                                                                                                                        &     \begin{tabular}[c]{@{}l@{}} \ion{Si}{IV} $\lambda$1393$\AA$$^{a}$ \\ \ion{Si}{IV} $\lambda$1403$\AA$ \\ \ion{C}{IV} $\lambda$1550$\AA$$^{a, b}$\end{tabular}   \\ \hline
        
         50\,000  & 4.2 & No detectable lines     &   No detectable lines    &    No detectable lines &  \begin{tabular}[c]{@{}l@{}}\ion{N}{V} $\lambda$1239$\AA$$^a$\\ \ion{O}{V} $\lambda$1371$\AA$$^a$\end{tabular}     \\
        \hline
         \begin{tabular}[c]{@{}c@{}} 50\,000\\ (WR-star) \end{tabular} & 4.0 & No detectable lines  &   
         
         No detectable lines   &    
         
         \begin{tabular}[c]{@{}l@{}}\ion{N}{IV} $\lambda$1488$\AA$\\ \ion{C}{IV} $\lambda$1550$\AA^a$\\ \ion{N}{IV} $\lambda$1720$\AA$ \end{tabular} &  
         
         \begin{tabular}[c]{@{}l@{}}\ion{N}{V} $\lambda$1239$\AA^a$\\ \ion{N}{IV} $\lambda$1488$\AA$\\ \ion{C}{IV} $\lambda$1550$\AA^a$  \end{tabular}    \\
        \hline 
        
        \end{tabular}

    \tablefoot{A few examples of mock spectra for particular $T_\mathrm{eff}$ and redshifts are displayed in Fig.~\ref{fig:Spectra6000-30000K}, Fig.~\ref{fig:Spectra4000-50000}, and Fig.~\ref{fig:SpectraWR}. The stellar atmosphere spectra based on models with $T_\mathrm{eff}\leq10\,000$ K use $[\mathrm{M/H}]=-1$, while the hotter stars use $[\mathrm{M/H}]=-0.845$.}
    \tablefoottext{a}{Also detectable at 27 AB mag.}
    \tablefoottext{a}{Part of a broader continuum feature.}
    \end{threeparttable}
\end{table*}

\subsection{Final mock spectra}
In the case of both JWST/NIRSpec and ELT/HARMONI mock spectra, noise is added to the final spectrum, based on either the output of the ETC (NIRSpec) or from the scaling relations above (HARMONI). For plotting purposes, these noisy spectra are also rebinned by averaging over a fixed number of spectral resolution elements, to make it easier to distinguish selected absorption features. These mock spectra do not include the noise that may result from subtraction of the local background. All high-redshift lensed stars detected to date have appeared on or close to gravitational arcs (the likely host galaxies of these stars). The light from these arcs typically dominate the background in the direct vicinity of the lensed stars and will inevitably contaminate the resulting spectra. This arc component should ideally be subtracted off, but since the surface brightness profiles of these components are not completely smooth, residuals are likely to linger in the observed spectra at some level. Because of this, we caution that the noise depicted at very low flux levels, for instance in saturated absorption features, may be underestimated.

\section{Detectable spectral features}
\label{sec:features}

\begin{figure*}[!htbp]

\includegraphics[scale=0.24]{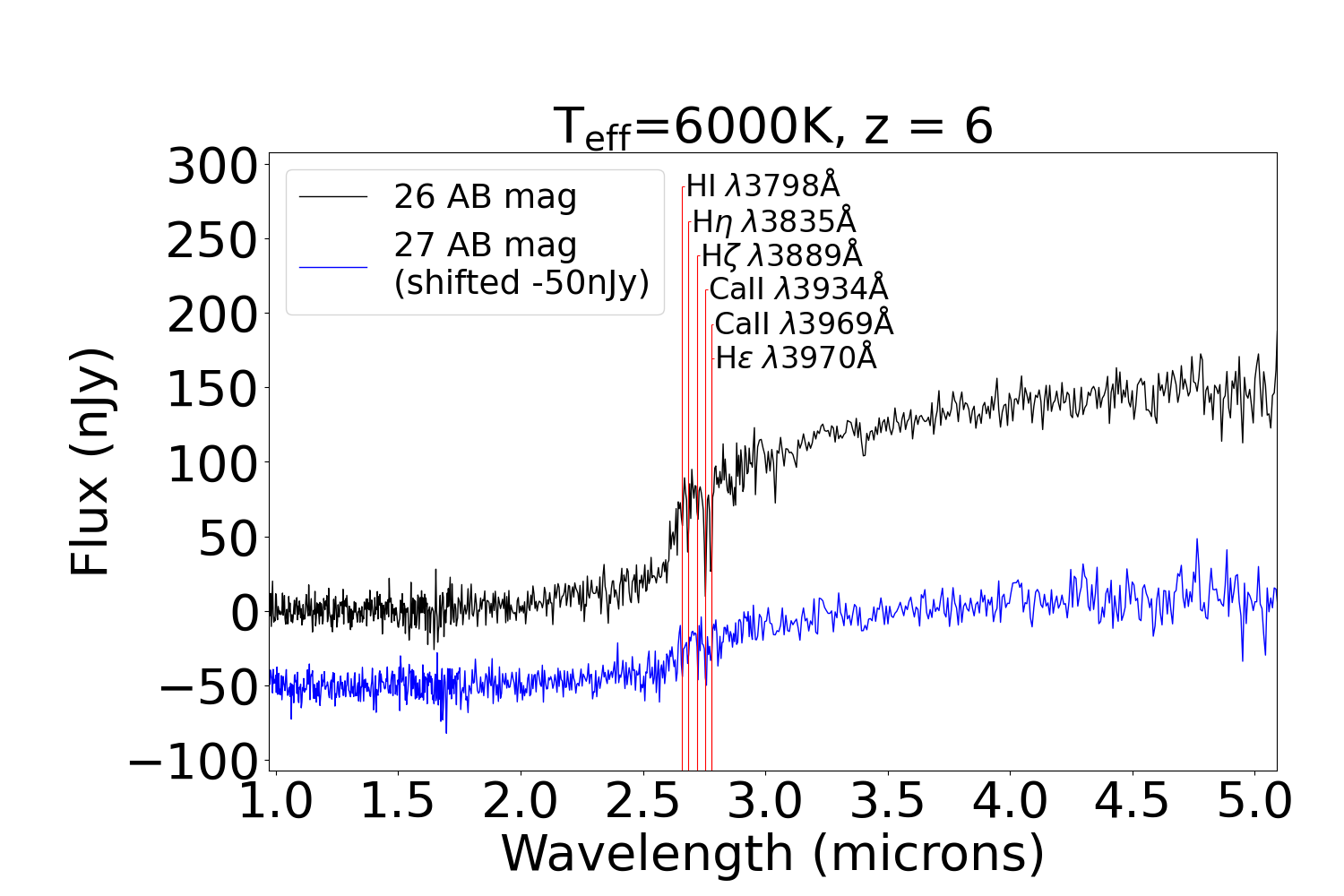}
\includegraphics[scale=0.24]{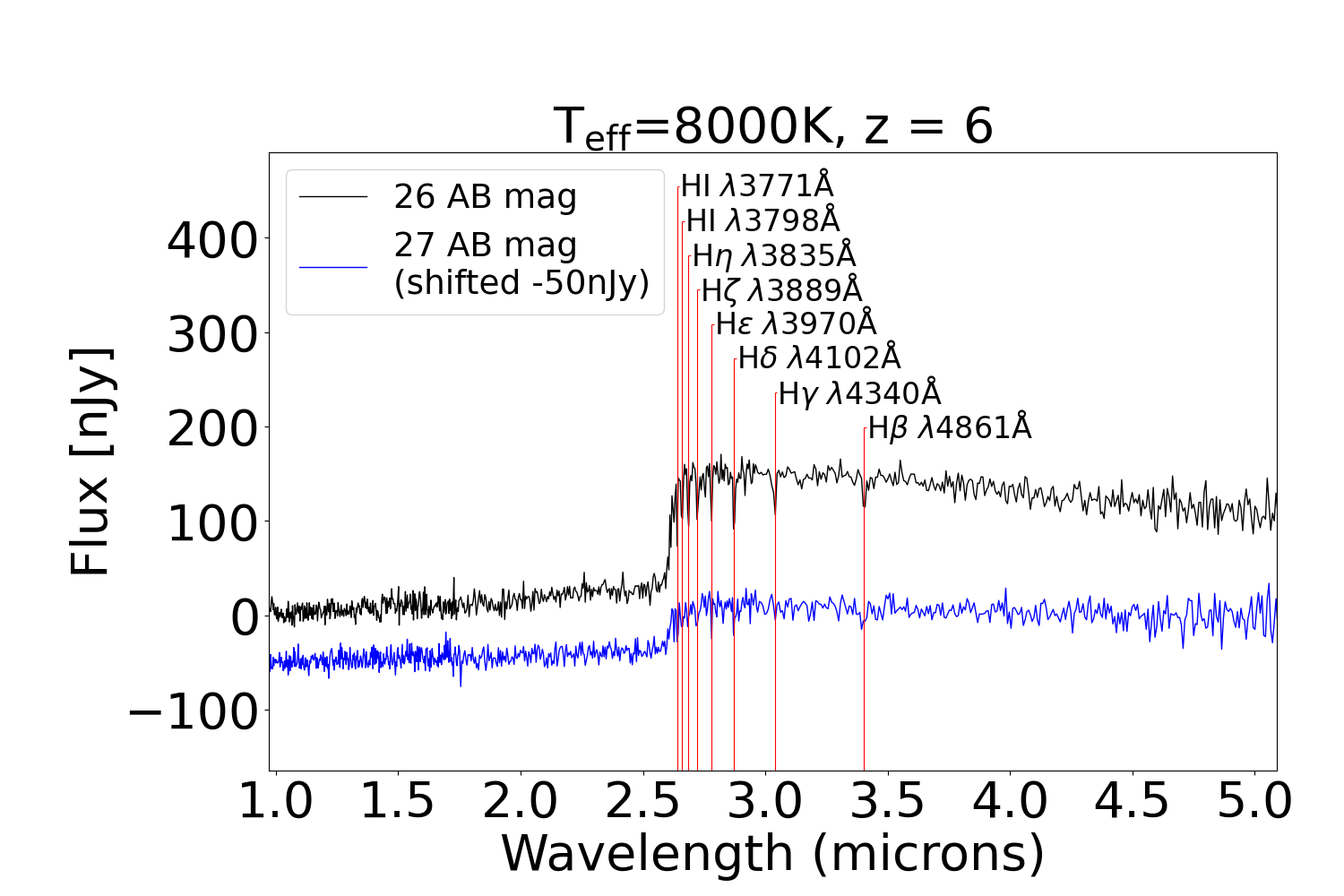}
\includegraphics[scale=0.24]{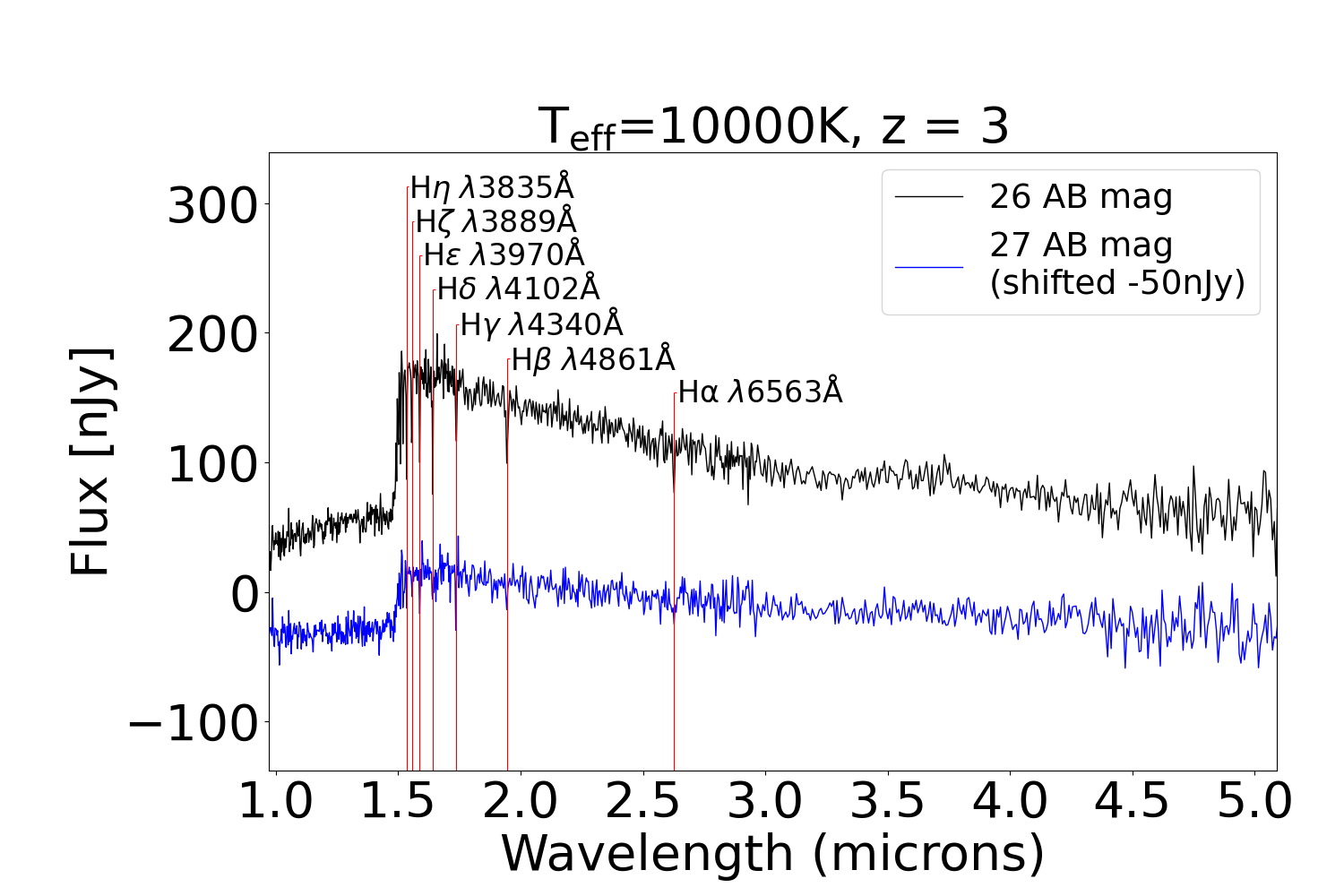}
\includegraphics[scale=0.24]{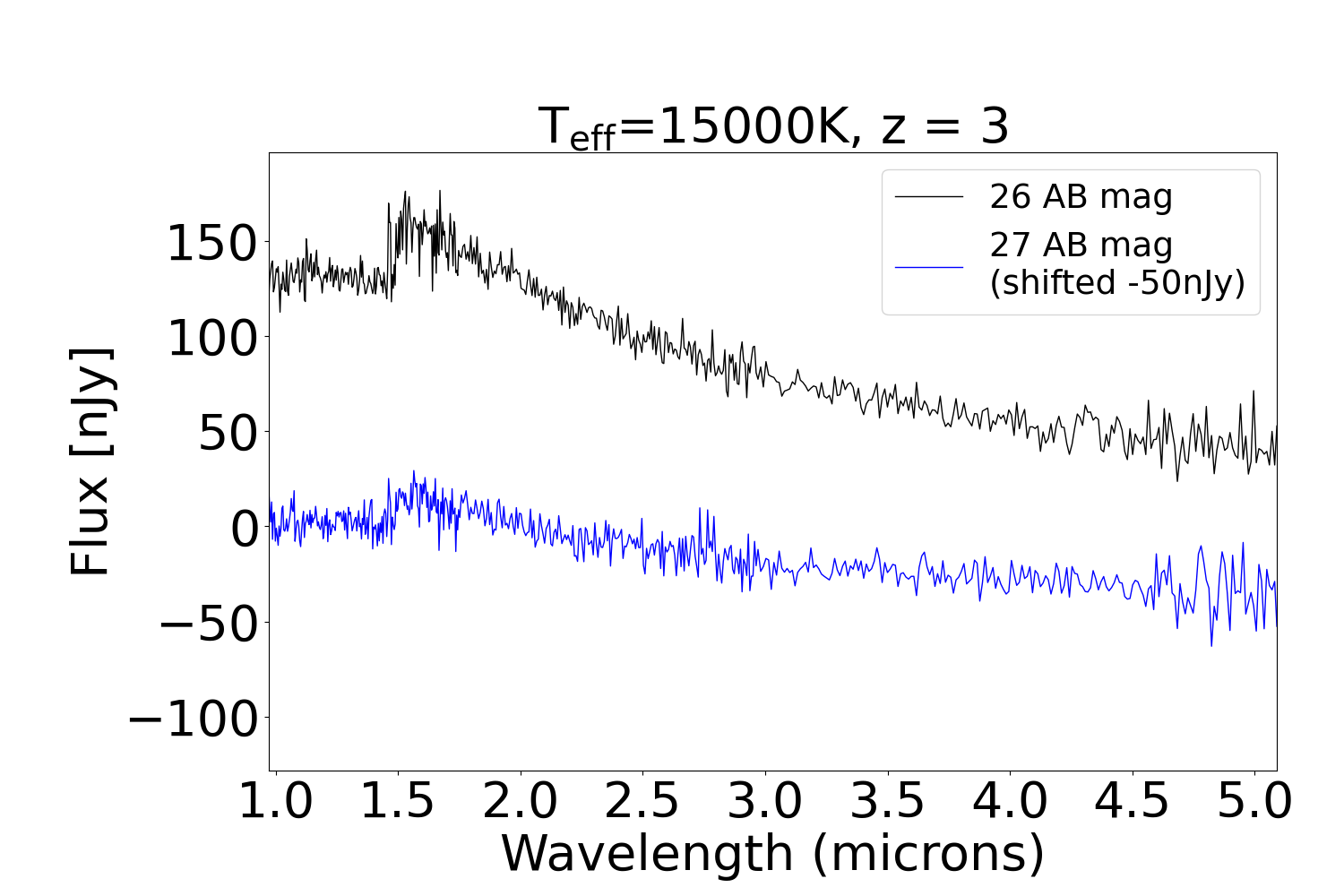}
\includegraphics[scale=0.24]{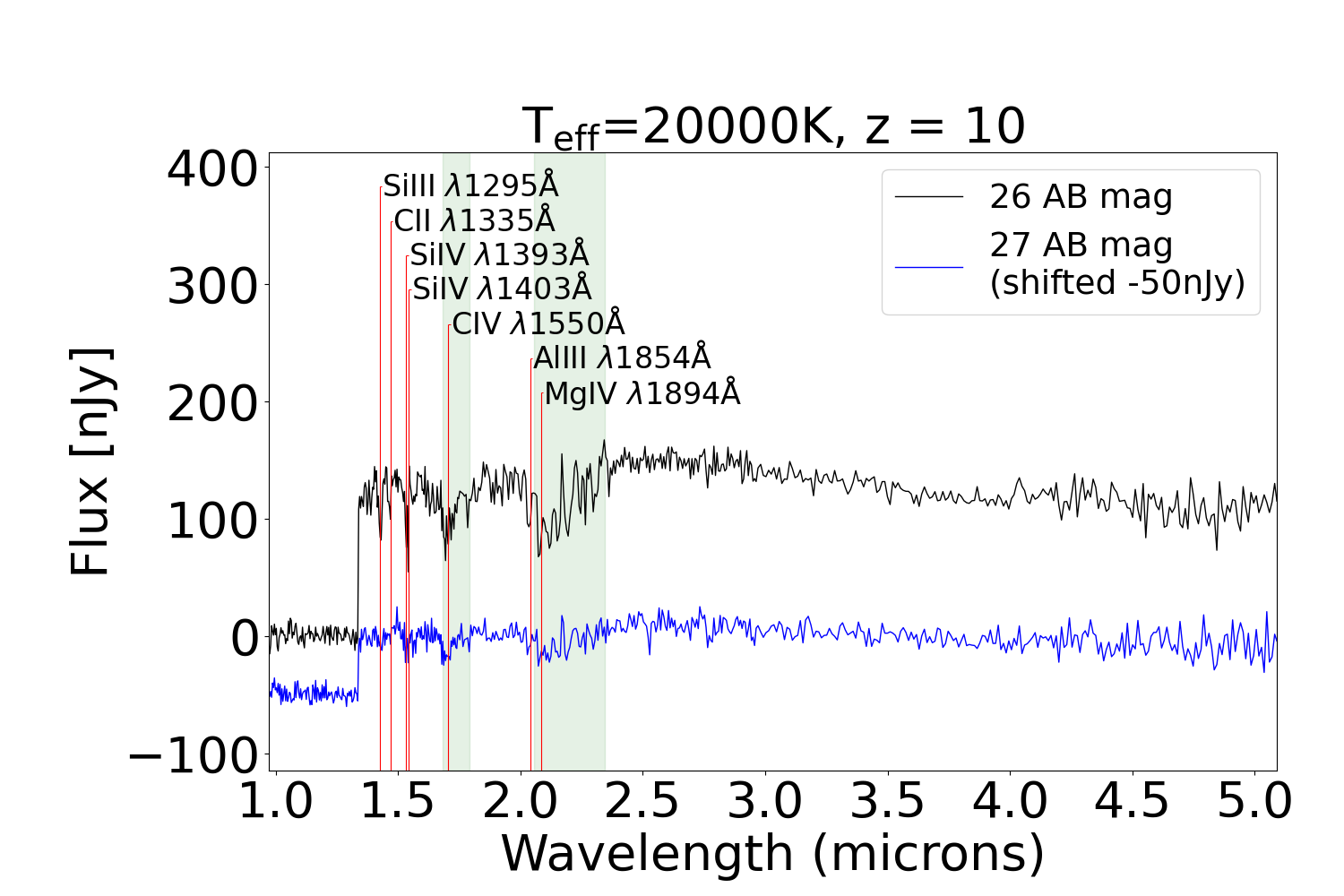}
\includegraphics[scale=0.24]{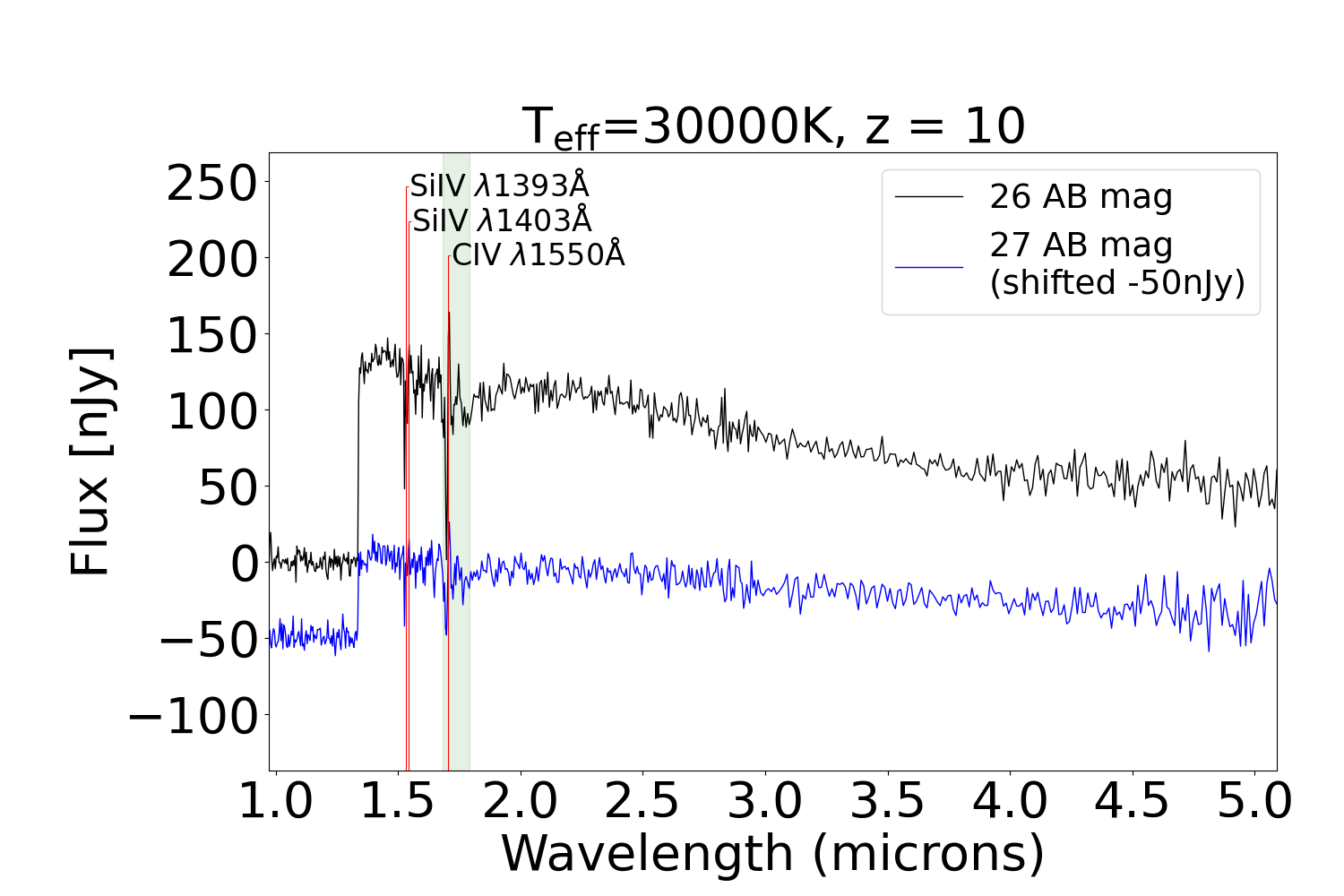}

\caption{Six examples of the noisy mock JWST/NIRSpec spectra, for six different effective temperatures at various redshifts. Each figure includes the spectra for both a peak magnitude of 26 (in black) and 27 (in blue) AB mag, where the 27 mag spectrum is shifted down by 50 nJy to avoid cluttering. The metal absorption lines included in Table~\ref{tab:MetalLines}, as well as hydrogen lines, are marked in red, indicating features detectable at 26 AB mag. The green regions mark broader features, within which at least one of the contributing lines has been identified. In all figures, except one, do detectable spectral lines appear. The exception, with $T_\mathrm{eff}=15000$ K and $z=3$, instead gives an example of a spectrum with no detections. The general conclusion from these figures is that stars with higher effective temperatures at higher redshifts result in a larger number of detectable metal lines, while cooler stars instead contribute with many hydrogen Balmer lines.}

\label{fig:Spectra6000-30000K}
\end{figure*}

In this study, a spectral feature is considered detectable if it can be identified with a significance of at least $5\sigma$, meaning spectral lines with larger errors are considered either unreliable detections or non-detections. For the strongest spectral line (\ion{C}{IV} $\lambda 1550$$\AA$ at 30\,000 K and 26 AB mag, actually part of a doublet together with \ion{C}{IV} $\lambda 1548$$\AA$, but here only referred to as the $\lambda 1550$$\AA$-line), this significance reaches as high as 18$\sigma$. This accuracy has been calculated using the equivalent width of the spectral lines, where a $5\sigma$ detection means that the equivalent width is larger than 5 times the standard deviation. The definition of the wavelength range over which the equivalent width is calculated is strongly connected to the resolution of the gratings used. Generally, the spectral lines encompass very few resolution elements, leading to the edges of the spectral lines being defined as being located between the last measurement point at continuum level and the first point within the line. The noise is included in the spectra by scaling a random number from a Gaussian distribution to the noise level at a particular wavelength point. The standard deviation of the calculated equivalent width was found via a Monte Carlo simulation, where the noise was allowed to change in each random sample.

\begin{figure*}[!htbp] 

\includegraphics[scale=0.24]{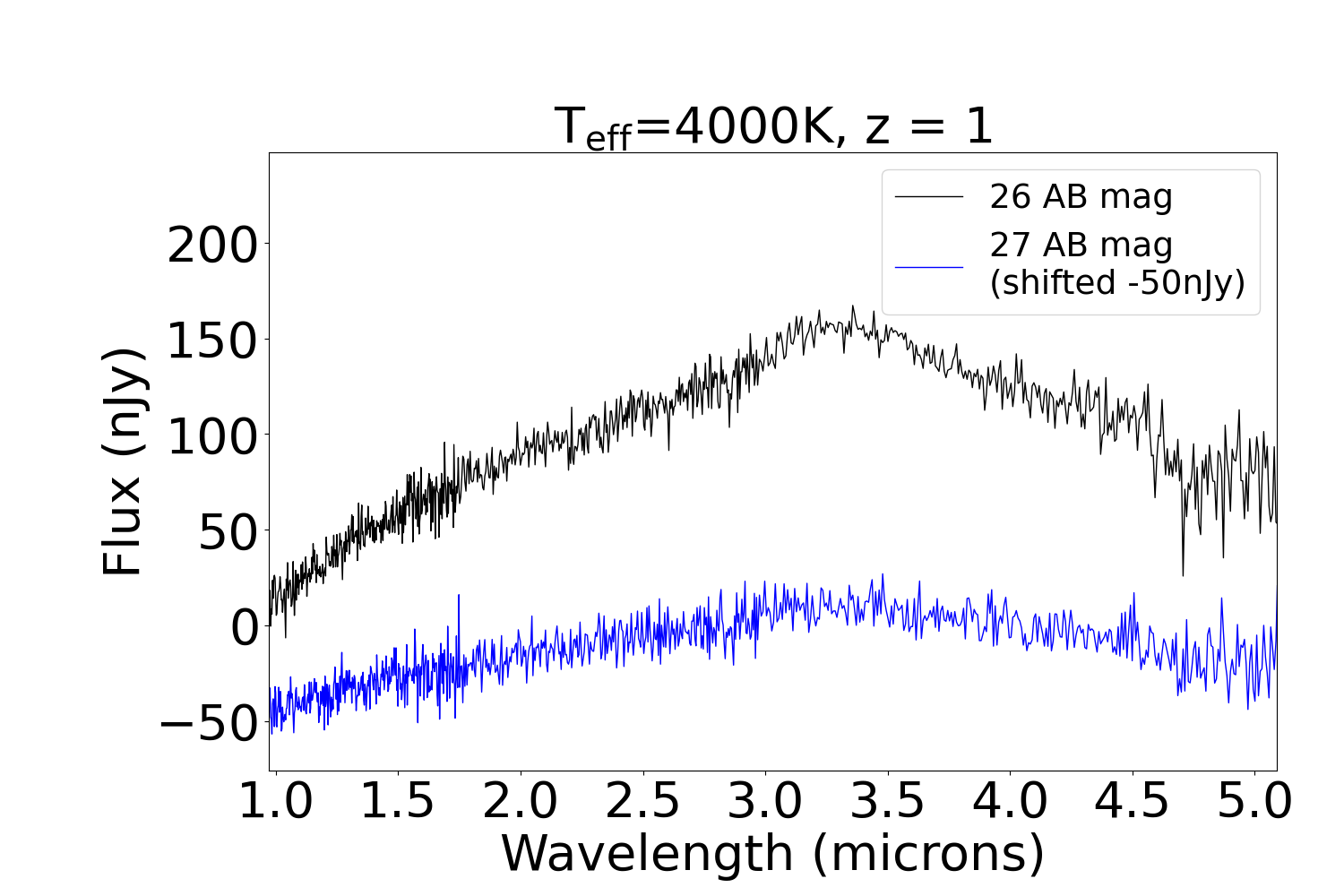}
\includegraphics[scale=0.24]{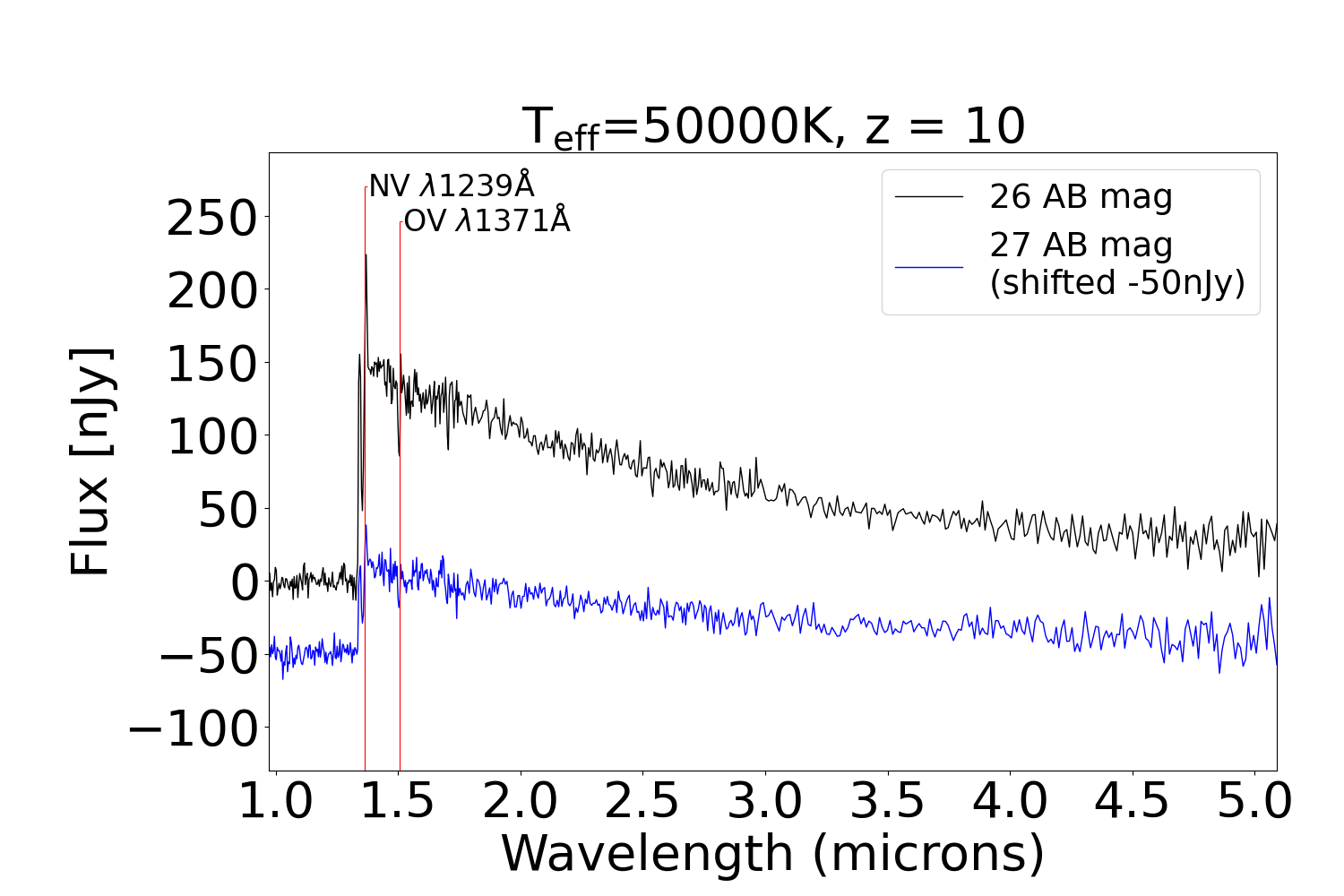}

\caption{Two examples of simulated stellar atmosphere spectra (using JWST/NIRSpec) for $T_{\mathrm{eff}}=4000$ K, $z=1$ on the left and $T_{\mathrm{eff}}=50\,000$ K, $z=10$ on the right. For the 4000K spectra, no spectral lines were detectable at any redshift at either 26 or 27 AB mag. Two metal lines can be observed at a peak magnitude of 26 mag for $T_{\mathrm{eff}}=50\,000$ K, and both of them can also be distinguished at 27 mag. These figures illustrate that few spectral lines are detectable at these temperatures, most apparently for 4000 K.}
\label{fig:Spectra4000-50000}

\end{figure*}

Table~\ref{tab:MetalLines} displays a summary of detectable metal lines using JWST/NIRSpec, assuming a peak magnitude of 26 AB mag, for the eight effective temperatures ($T_\mathrm{eff}=4000$, 6000, 8000, 10\,000, 15\,000, 20\,000, 30\,000, and 50\,000 K) and all four redshifts ($z=1$, 3, 6, and 10), with the addition of a Wolf-Rayet (WR) star at 50\,000 K. A subset of the strongest lines also appear detectable at 27 AB magnitudes. An additional type of spectral feature is apparent in certain spectra, namely broad dips in the continuum caused by the blending of multiple lines. They are caused by Fe-forests, blends of a large number of iron lines in very close proximity to each other, one of which is predominantly caused by \ion{Fe}{IV} and appears at a rest wavelength range of $\sim1530$--$1630$ $\AA$. The second one is primarily caused by \ion{Fe}{III} and affects the continuum around $\sim1870$--$2130$ $\AA$. A table including the hydrogen and helium lines detectable at 26 AB mag can be found in the Appendix, Table~\ref{tab:HLines}.

\begin{figure}

\includegraphics[scale=0.24]{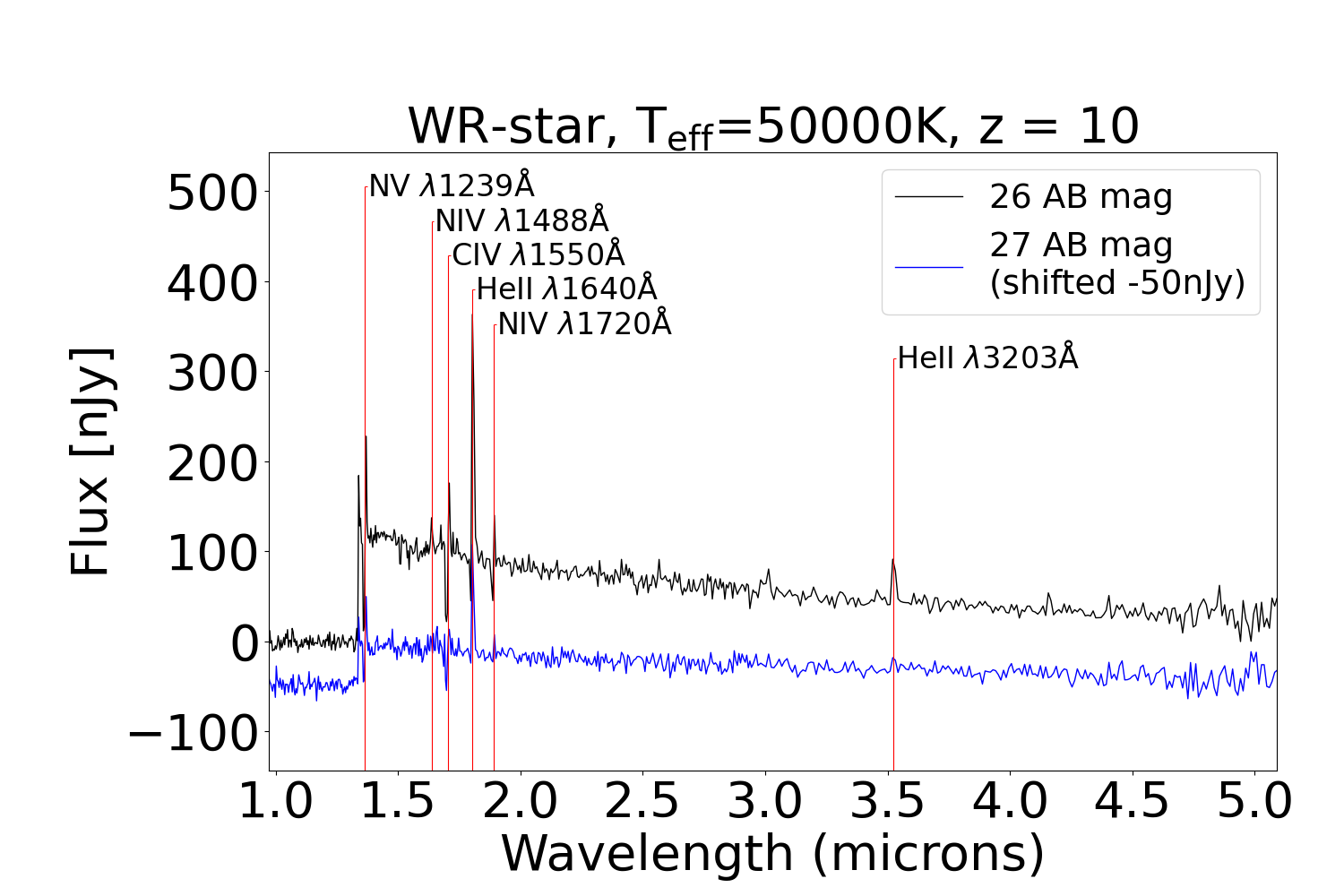}

\caption{JWST/NIRSpec mock spectra of a WR star at $T_\mathrm{eff}=50\,000$ K and $z=3$. Compared to the spectra of a 50\,000 K-star in Fig.~\ref{fig:Spectra4000-50000}, the number of detectable lines is more numerous. Most significantly, strong helium emission lines appear.}
\label{fig:SpectraWR}

\end{figure}

\subsection{Examples of JWST mock spectra}
Figs.~\ref{fig:Spectra6000-30000K}, \ref{fig:Spectra4000-50000}, and \ref{fig:SpectraWR} depict a few examples of the NIRSpec mock spectra, in which spectral lines have been identified. Each individual figure includes two spectra, normalised to the peak magnitudes $m_\mathrm{AB}=26$ and $m_\mathrm{AB}=27$, where the $m_\mathrm{AB}=27$ spectrum is shifted to avoid overlap. Whereas only metal lines are included in Table~\ref{tab:MetalLines}, all spectral lines are marked in the figures to further clarify what lines are considered detectable. Fig.~\ref{fig:Spectra6000-30000K} displays one plot each for the effective temperatures between $6000$--$30\,000$ K, at various redshifts. These six figures provide a good summary of all generated mock spectra for these temperatures. In most cases, the same lines are detectable for the same temperatures at different redshifts (as long as they lie within the observable spectral range of JWST), meaning that one example can represent multiple redshifts. 

Very few metal lines are detectable in the mock spectra for the cooler stars, $T_\mathrm{eff}\leq 10\,000$ K, and 6000 K is the sole effective temperature for which these stars exhibit any clear, detectable metal lines that are not strongly blended by hydrogen lines. The upper left figure in Fig.~\ref{fig:Spectra6000-30000K} depicts $T_\mathrm{eff}=6000$ K spectra, where the Ca-H and Ca-K lines, at $\lambda=3969$ $\AA$ and $\lambda=3934$ $\AA$ respectively are easily detectable. These lines peak in strength around 6000 K, and at 8000 K and 10\,000 K (upper right and middle left figures) Ca-K is no longer detectable and Ca-H is blended with the H$\varepsilon$ line. In these cases, the Ca-H line is strongly dominated by the hydrogen line, meaning that no metal lines are actually detectable at 8000 K and 10\,000 K. The upper right figure illustrates how this blended line becomes dominated by the hydrogen component at 8000 K, as it shows a figure with the same redshift as the upper left one. The middle left figure also presents a situation with only hydrogen lines, similar to the upper right one. The log-ratio ($\log(W_{\lambda, H}/W_{\lambda, Ca})$) of the respective equivalent widths of the Ca-H line and H$\varepsilon$ changes significantly when the temperature increases. For 6000 K, the calcium line is stronger, leading to $\log(W_{\lambda, H}/W_{\lambda, Ca})\sim-0.3$, while for 8000 K and 10\,000 K the hydrogen line is stronger with $\log(W_{\lambda, H}/W_{\lambda, Ca})\sim0.7$ and $\log(W_{\lambda, H}/W_{\lambda, Ca})\sim1.4$ respectively.

As mentioned, the 6000 K case is the only one among the lower effective temperatures that feature detectable metal lines. Except for the Ca-K and -H lines, two lines of the Ca infrared triplet (\ion{Ca}{II} $\lambda\lambda$ 8542$\AA$, 8662$\AA$) are discernible at $z=1$, but only detectable at $\sim4\sigma$, and therefore not included in Table~\ref{tab:MetalLines}. At higher redshifts ($z>1$), the Ca-H and Ca-K lines enter the NIRSpec wavelength range. These are among the strongest spectral lines in the solar spectrum, at a temperature close to $T_\mathrm{eff}=6000$ K.

Three of the figures in Fig.~\ref{fig:Spectra6000-30000K} show spectra for hot stars, with $T_\mathrm{eff}\geq15000$ K. The middle right figure in Fig.~\ref{fig:Spectra6000-30000K} illustrates spectra completely dominated by noise, as no spectral lines are detectable at either of the peak magnitudes. As can be seen in Table~\ref{tab:MetalLines}, very few metal lines are detectable at $z=1$ and $z=3$ at any temperature, in particular for the hotter stars. The bottom left figure instead demonstrates a spectrum with a large number of detectable spectral lines, and very prominent Fe-forests, a situation which only occurs at high temperatures and in the rest-frame UV-range. Within and between the Fe-forests, a great number of individual lines could be considered detectable, but for the sake of clarity, only the most prominent ones are marked. The figure in the bottom right includes a few of the lines detected at the highest significance, some of them also detectable at 27 AB mag. These lines exhibit P Cygni profiles, indicating that they are dominated by stellar winds.

Table~\ref{tab:MetalLines} clearly shows that high redshifts are necessary for metal lines in spectra of high temperature stars ($T_\mathrm{eff}\gtrsim15\,000$ K) to be detectable with NIRSpec. In none of the spectra for these stars do any metal lines appear at a redshift lower than six. This is due to the fact that every single prominent line for these high $T_\mathrm{eff}$ stars lies in the rest-frame UV-range (the only exception here being the WR-star). Hot stars are brighter at shorter wavelengths, increasing the continuum level towards the blue and UV (see high-$T_\mathrm{eff}$ examples in Figs.~\ref{fig:Spectra6000-30000K} and \ref{fig:Spectra4000-50000}), resulting in a higher S/N and making the spectral lines at these wavelengths more easily detectable. Also, most spectral lines in the spectra of hot stars appear in the UV-range. The largest number of detectable metal lines is present at $T_\mathrm{eff}\gtrsim15\,000$--20\,000 K, and the strongest ones at $T_\mathrm{eff}\gtrsim30\,000$ K. Hot lensed stars are therefore interesting to observe with JWST/NIRSpec, but a redshift of $z\gtrsim3$ is needed to detect spectral features. The exception is the WR-star, where multiple spectral lines, in particular helium lines, are detectable at optical rest wavelengths.

An interesting note is that the strong \ion{C}{IV}-feature seen in the 30\,000 K-spectra has already been detected in spectra of a lensed star, namely that of Godzilla \citep{Diego22}. NIRSpec integral field spectroscopy of Godzilla has also revealed multiple emission lines at optical rest-wavelengths \citep{Choe24}. The nature of Godzilla has been debated though, as \cite{Pascale24} suggest that it is a lensed super star cluster as opposed to a lensed star.

Two of the effective temperatures investigated here, 4000 K and 50\,000 K, are less likely to be observed as lensed stars, at least at high redshifts. At 4000 K, the spectrum is mostly redshifted out of JWST's observable wavelength range at high redshifts, and there is currently no empirical evidence for resolved lensed stars at 50\,000 K at any redshift. This might also be connected to the degeneracy between $T_\mathrm{eff}$ and dust extinction, as some very hot stars with high levels of dust extinction might be mistaken for cooler stars. Also, very few metal lines are detectable at these temperatures. Fig.~\ref{fig:Spectra4000-50000} shows one figure for $T_\mathrm{eff}=4000$ K and one for $T_\mathrm{eff}=50\,000$ K. No spectral lines are detectable for 4000 K, at any redshift, exemplified by the figure at $z=1$. For 50\,000 K, only metal lines at short rest-frame UV-wavelengths are detectable, first showing up at $z=10$. The strongest of these lines, the \ion{N}{V} $\lambda\lambda 1239, 1243$$\AA$-doublet (referred to as the \ion{N}{V} $\lambda 1239$$\AA$-line), lies very close to the Lyman-$\alpha$ break at a rest wavelength of $1216$ $\AA$. This break is caused by the neutral intergalactic medium absorbing photons with wavelengths shorter than $1216$ $\AA$, and is visualised by a significant drop in flux at this particular wavelength. Our treatment of the Lyman-$\alpha$ break is admittedly not entirely realistic, as the Lyman-$\alpha$ damping wing can extend to longer wavelengths than depicted here. The \ion{N}{V}-line is then likely to, at least partly, disappear into the Lyman-$\alpha$ break, which would make it harder to observe.

\begin{figure*}
    \includegraphics[scale=0.35]{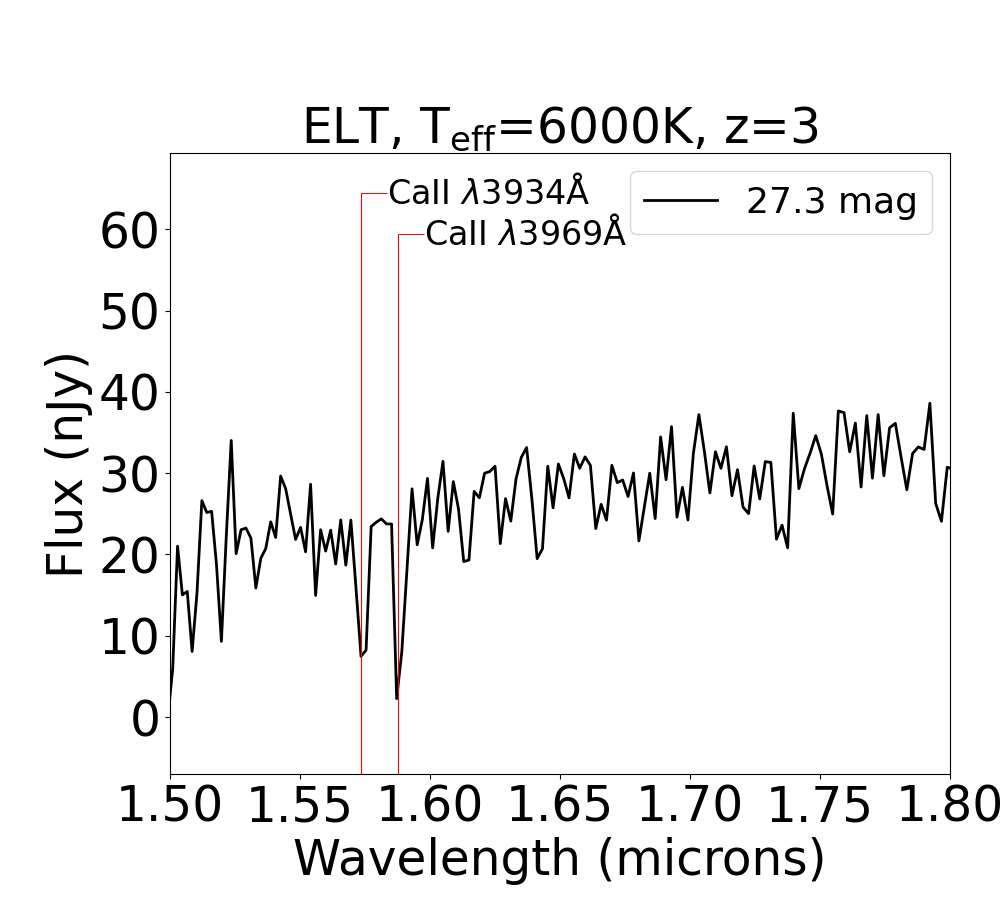}
    \includegraphics[scale=0.35]{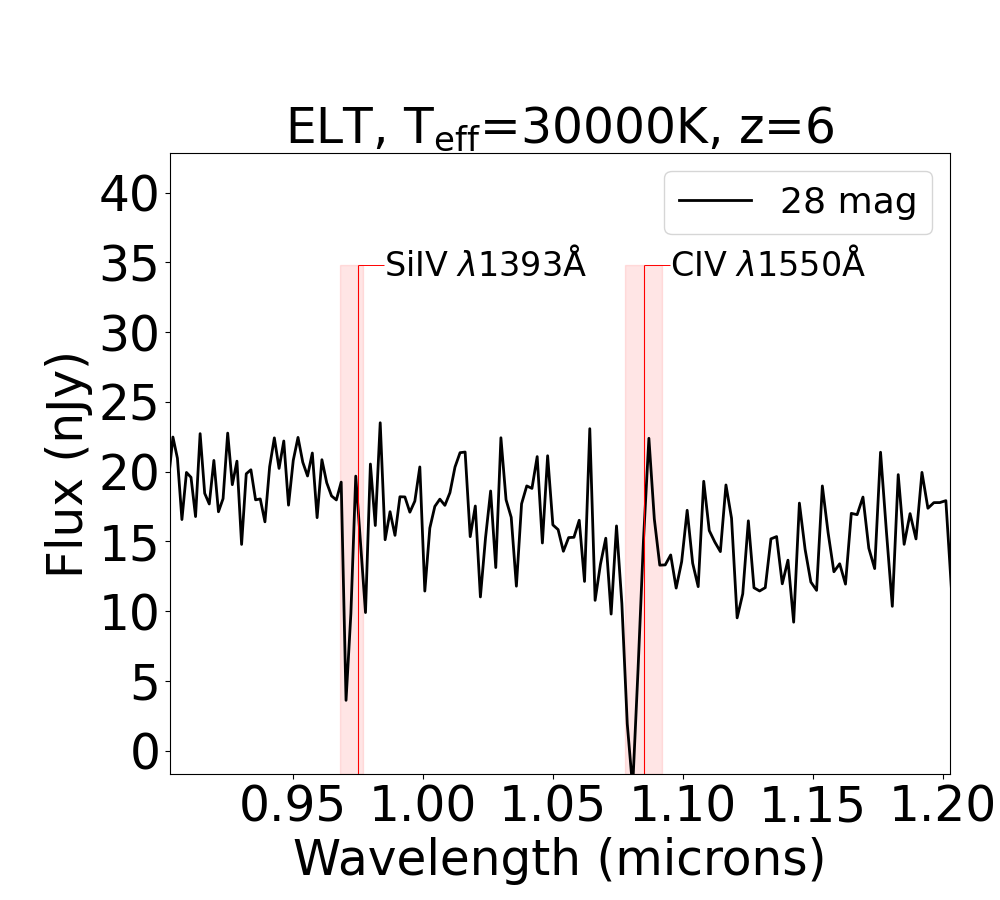}
    \caption{Mock ELT/HARMONI spectra focusing on the strongest lines detectable in the mock JWST/NIRSpec spectra. Both figures illustrate that features will be distinguishable in ELT/HARMONI spectra for fainter stars, and with shorter exposure times (5 hours for the left figure and 10 hours for the right one) than for JWST/NIRSpec. To clarify which lines are marked in the right figure, the areas encompassing the P Cygni profiles are filled.}

    \label{fig:ELT_figures}
\end{figure*}

These results indicate that no spectral lines will be detectable with NIRSpec for lensed stars with an effective temperature around 4000 K. \cite{Zackrisson23} found that at $z<6$, such lensed stars will be favourably detected, but according to this study no features will be detectable in spectroscopic observations of them at apparent magnitudes $\geq$ 26 AB mag. On the other hand, a few spectral lines are detectable at 50\,000 K.

Fig.~\ref{fig:SpectraWR} displays spectra for a hydrogen free WR star in the nitrogen subclass at $T_\mathrm{eff}=50\,000$ K and $z=10$. The star is also included in Tables~\ref{tab:MetalLines} and~\ref{tab:HLines}. Compared to the 50\,000 K-star in Fig.~\ref{fig:Spectra4000-50000}, the number of detectable lines for the WR-star is significantly increased. A few new metal lines have appeared, but also a number of very strong helium lines in emission, characteristic for WR-stars (see table~\ref{tab:HLines}). Even at low redshifts, $z\leq3$, do strong helium lines appear. Many of them are detectable at 27 AB mag, and the strongest one, \ion{He}{II} $\lambda1640$ $\AA$, also at 28 AB mag. It is unclear whether WR-stars are likely to appear in samples of lensed stars, however, and as mentioned, lensed stars with these very high temperatures have not yet been observed.

\subsection{Impact of redshift on line detectability}
The redshift influences whether a spectral line is detectable with NIRSpec or not, depending on if the line, with the added redshift, enters the 1--5 micron range of JWST. But in a few cases, for a given effective temperature, a line may lie within this range for multiple redshifts, yet still only be detectable at some subset of them. This implies that the redshift affects the detectability of the lines in additional ways. The grating/filter pairs we use have wavelength-dependent resolution. Since the observed wavelength of a line changes with redshift, the resolution of the spectrograph for a particular spectral line also depends on the redshift. Generally, the resolution increases with wavelength (except at the transition between gratings), meaning that a higher redshift leads to a higher resolution, as the observed wavelength of a particular line has increased. This higher resolution could then impact the detectability of spectral features.

\subsection{ELT mock spectra}

Fig.~\ref{fig:ELT_figures} shows two examples of mock spectra generated for ELT/HARMONI, focusing on particular spectral features. The spectrum to the left is normalised to the calibration magnitude 27.3 AB mag, and investigates the Ca-H and Ca-K lines. The effective temperature is $T_\mathrm{eff}=6000$ K and the exposure time $t_\mathrm{exp}=5$ hours. These lines are not detectable at this magnitude for JWST/NIRSpec, but are easily discernible in this spectrum simulating an observation with ELT. The implication of this result is that ELT will be a more suitable telescope for spectroscopy of lensed stars, in this particular wavelength range. The right figure also reinforces this for a different wavelength range, where an analysis is done regarding how faint a star can be for the ELT to still be able to discern spectral features within a reasonable exposure time. For this, the strongest of the identified spectral lines are used, \ion{C}{IV} $\lambda1550$$\AA$ and the \ion{Si}{IV} $\lambda\lambda 1393, 1403$$\AA$-doublet at 30\,000 K and $z=6$. With the ELT, two of these lines can be detected in a star as faint as 28 AB mag, with an exposure time of $t_\mathrm{exp}=10$ hours. Only one of the Si lines is marked, since the second is too weak to be considered detectable, with a detection significance of $\sim3.6\sigma$.

Although ELT/HARMONI seems like the better option for spectroscopy of lensed stars, it is important to also note its disadvantages. The spectral range in the near-infrared for ELT is much smaller than for JWST, $\lambda \sim 0.811$ -- 2.450 $\mu$m instead of $\lambda\sim0.97$--5.1 $\mu$m. Additionally, certain wavelengths of near-infrared light are absorbed in the atmosphere, creating gaps throughout the HARMONI spectral range. 

\section{Discussion}
\label{sec:discussion}

\begin{figure*}
    \includegraphics[scale=0.35]{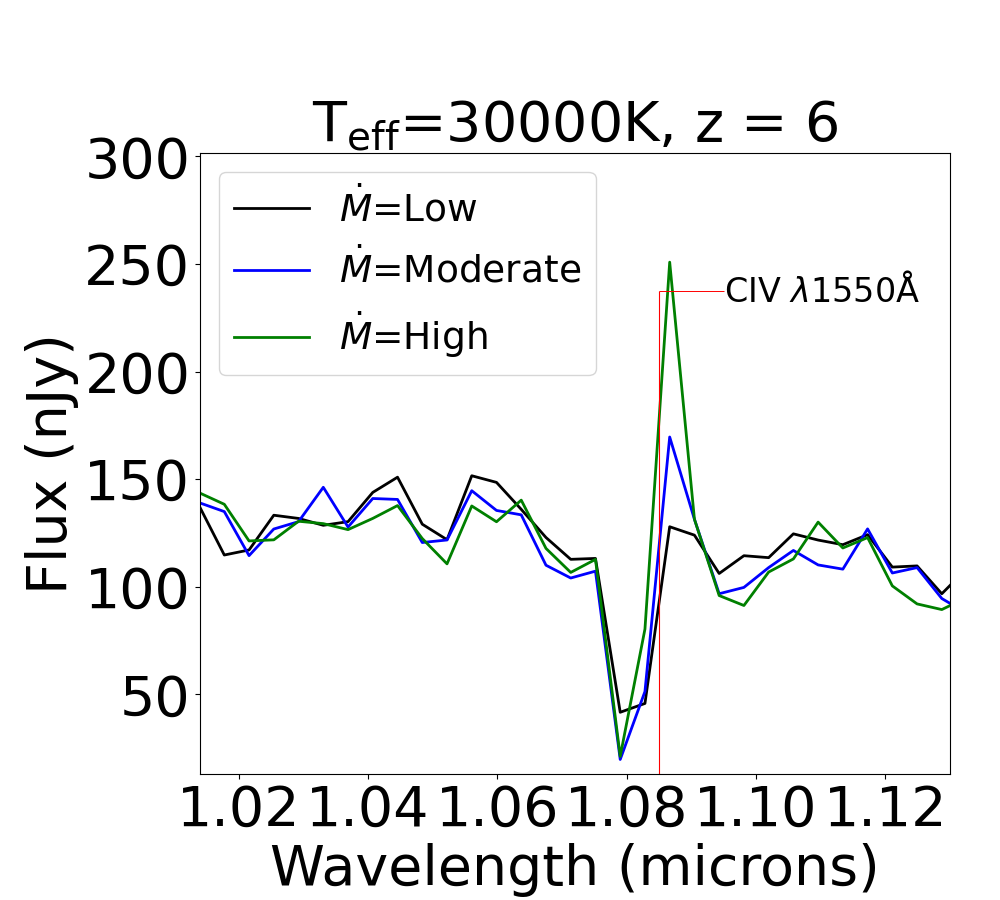}
    \includegraphics[scale=0.35]{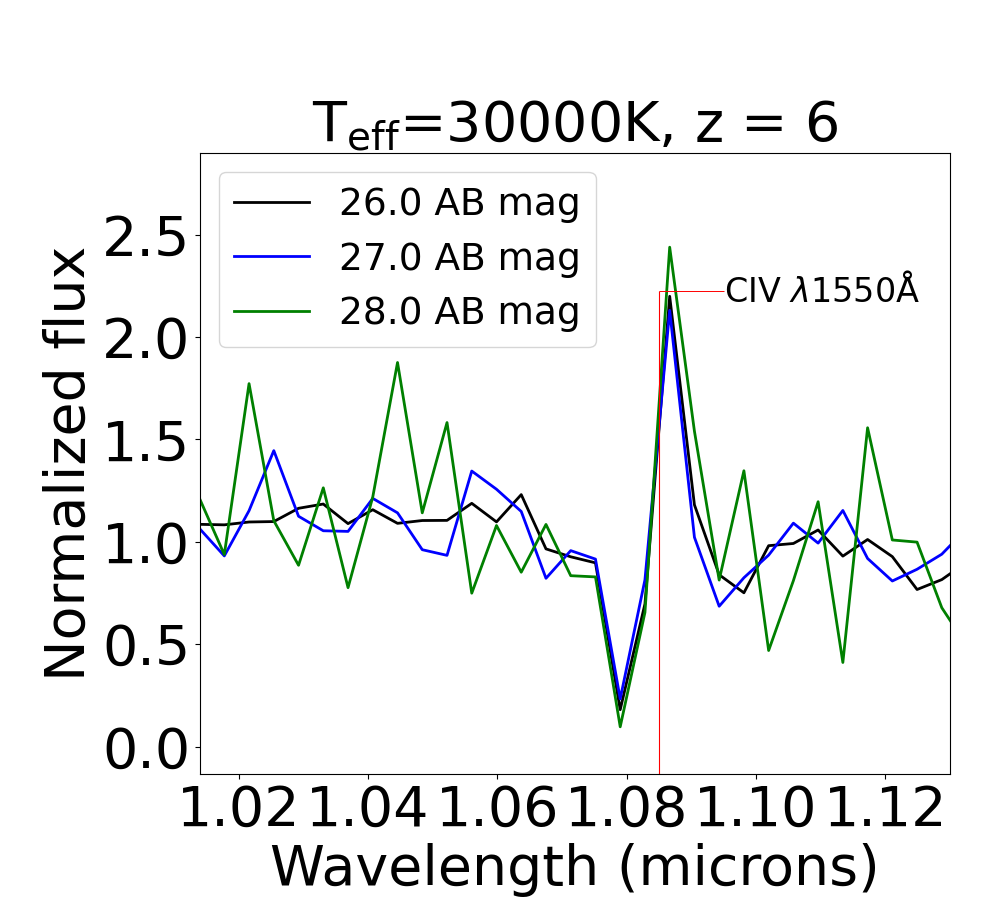}
    \caption{JWST/NIRSpec mock spectra illustrating how the mass loss rate affects the detectability of a spectral line, in particular \ion{C}{IV} $\lambda$1550$\AA$. The figures depict the spectra for a star with $T_\mathrm{eff}=30\,000$K and a redshift of $z=6$. The left figure compares different mass loss rates and illustrates its effect on the strength of the emission part of the line, while the right figure shows how the spectral line is consistently detectable up to 28 AB mag if a high mass loss rate is used.}

    \label{fig:Wind_30000K}
\end{figure*}

\subsection{Use of detected features}
\label{Sec:use-of-features}

The spectral lines could possibly provide information about physical parameters of the lensed stars, which will be investigated more in detail in Sects.~\ref{winds} and~\ref{abundance}. Due to P Cygni profiles being caused by stellar wind, lines with strong P Cygni absorption profiles contain information about the mass-loss, for example about $\dot{M}$ and $v_\infty$. $\dot{M}$ is the mass-loss rate of the star, and $v_\infty$ the terminal wind velocity, the velocity the stellar wind achieves very far from the surface of the star and the maximum velocity of the wind. When the emission components are detectable, the information becomes more robust, and more than just an upper limit can be determined. In $T_\mathrm{eff}=30\,000$ K-stars, the \ion{C}{IV} $\lambda\lambda1548, 1550$$\AA$ and \ion{Si}{IV} $\lambda\lambda$1393, 1403$\AA$ doublets are the strongest P Cygni profiles, and in $T_\mathrm{eff}=50\,000$ K-stars the \ion{N}{V} $\lambda\lambda 1239, 1243$$\AA$-doublet. At lower temperatures these features are weaker. Terminal wind speed measurements can be made from any P-Cygni line but are more accurate the closer the line is to saturation. Information about effective temperature and spectral type is limited, but rough estimations could be made with detections of photospheric profiles. Photospheric CNO-lines can, additionally, provide abundance diagnostics.

Although these spectral lines have their uses in measuring stellar parameters, the spectral resolution is still quite low, which will introduce uncertainties into the measurements. It is generally challenging to determine physical parameters from low-S/N detections of single lines, due to the inherent limitations of the observations. For example when estimating $v_\infty$, the parameter is directly read from the spectrum, as the difference between the furthest extent of the absorption valley in the P Cygni profile and the rest wavelength of the line. A low resolution, and high noise volume, might shift these points. By shifting the points, the measured wavelengths used to calculate $v_\infty$ are likely to change as well, leading to a less accurate calculation of the parameter.

\subsection{Impact of mass loss rate}
\label{winds}

Stellar winds are included in the stellar atmosphere modelling for the PoWR grids, here used for mock spectra of lensed stars with $T_\mathrm{eff}\geq15\,000$ K. For a metallicity equal to that of the SMC, grids with three different mass loss rates are available, low ($\log\dot{M}\left[\mathrm{M_\odot yr^{-1}}\right]\sim-7.7$), moderate ($\log\dot{M}\left[\mathrm{M_\odot yr^{-1}}\right]\sim-6.7$), and high ($\log\dot{M}\left[\mathrm{M_\odot yr^{-1}}\right]\sim-5.7$), where a moderate mass loss rate was used for previous results. While changing the mass loss rate of the models, and keeping the effective temperature fixed, the terminal wind velocity is kept constant. A comparison between the three mass loss rates, in Fig.~\ref{fig:Wind_30000K}, reveals its effect on the detectability of spectral lines, here showing the P Cygni profile of \ion{C}{IV} $\lambda1550$$\AA$ at $T_\mathrm{eff}=30\,000$ K. The left figure in Fig.~\ref{fig:Wind_30000K}, JWST/NIRSpec mock spectra normalised to a peak magnitude of 26 AB mag, directly compares the three levels of $\dot{M}$. The absorption component of the P Cygni profile appears weakly dependent on the mass loss rate in this case, as it is nearly identical for high and moderate $\dot{M}$. There is a slight decrease in strength for low $\dot{M}$, though, still indicating a detectable dependence on the mass loss rate. The reason for the shape being so similar for moderate and high $\dot{M}$ is likely that the line becomes saturated already at a moderate mass loss rate. The emission part, on the other hand, displays a large variation between different values of $\dot{M}$. The emission line clearly grows stronger with an increased mass loss rate, and the right figure illustrates how a high mass loss rate allows for this component to be detectable also for very faint stars ($m_\mathrm{AB}=28$). Although the noise increases at fainter magnitudes, the emission part of the P Cygni profile remains detectable. In the case of a moderate mass loss rate, the emission component gets lost in the noise at 28 AB magnitudes.

These results imply that spectroscopy of very faint lensed stars (at 28 AB mag) might result in a detectable spectral line with $\log\dot{M}\left[\mathrm{M_\odot yr^{-1}}\right]\sim-5.7$, which is not the case for a lower mass loss rate. Consequently, a detection of the \ion{C}{IV} $\lambda1550$$\AA$ line in emission at 28 AB mag requires the mass loss rate of the observed star to be high, since the line would not be detectable otherwise. For even higher values of $\dot{M}$, the P Cygni profiles are expected to only become stronger. In the cases where the absorption component is already saturated at these mass loss rates, mainly the emission component will be affected by a change in $\dot{M}$.

We caution, however, that the presence of an interstellar \ion{C}{IV} $\lambda1550$$\AA$ line in principle could contaminate measurements of this type. This line has been seen in metal-poor galaxies at both low and high redshifts \citep[e.g][]{Izotov24,Topping24,Castellano24}, and while nebular emission from the ambient interstellar medium of the host galaxy of a lensed star can sometimes be subtracted off \citep{Furtak24}, a parsec-scale nebular region in the vicinity of the lensed star could experience significant magnifications of up to $\approx 1000$ \citep[e.g.][]{Zackrisson23} and may contribute lines to its spectrum that are not easily corrected for.

Only for the hotter stars are stellar winds included in the stellar atmosphere modelling. It might be possible, though, that a small number of the most luminous cool stars drive strong stellar winds as well, due to the proximity to the Eddington limit. This could increase the number of detectable spectral lines for these cooler stars, due to the inclusion of wind-driven lines.

\begin{figure}
    \includegraphics[scale=0.35]{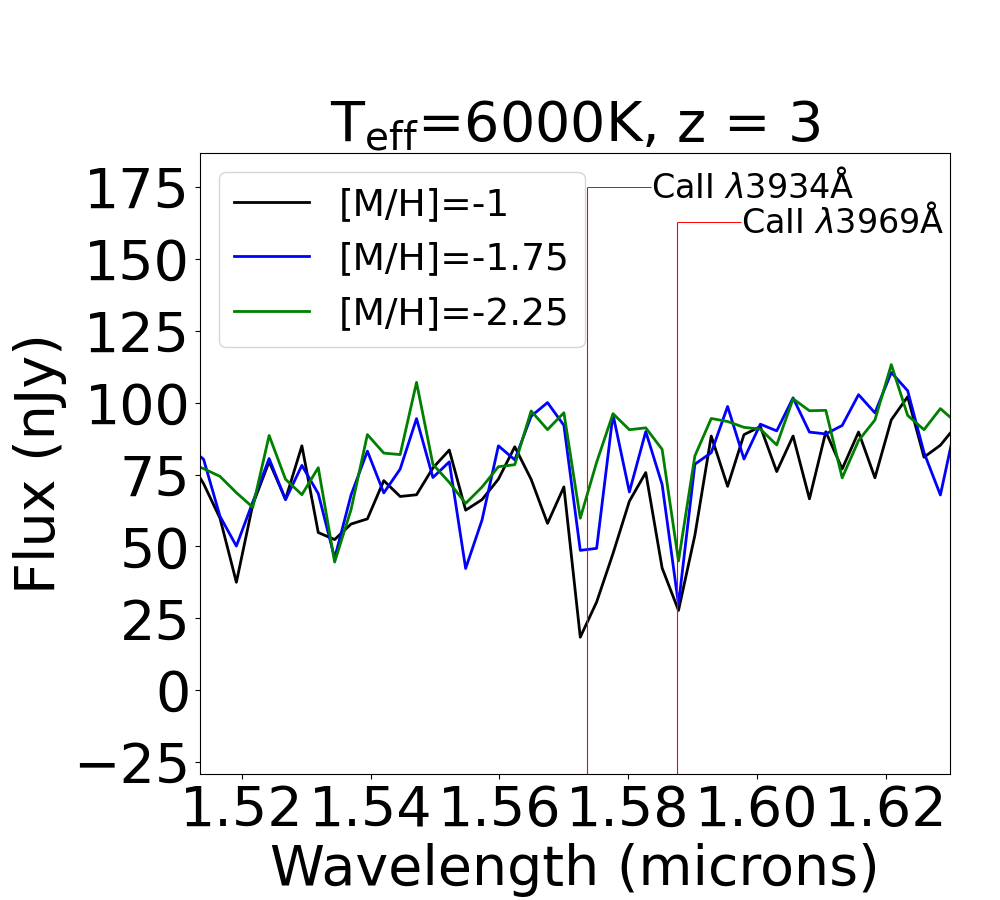}
    \caption{JWST/NIRSpec mock spectra depicting the Ca-H (at 3969$\AA$) and Ca-K (at 3934$\AA$) lines at $T_\mathrm{eff}=6000$K and $z=3$, comparing different metallicities, in particular studying the effect of the Ca-abundance. Ca-K remains detectable at metallicities as low as $[\mathrm{M/H}]=-2.25$ (at $\sim4\sigma$), and displays a Ca-abundance sensitivity that may allow it to be determined to within 0.75 dex. Ca-H is blended with a hydrogen line, and thus not as affected by the changing Ca-abundance.}
    \label{fig:Metallicity_6000K}
\end{figure}

\subsection{Impact of Ca-abundance and metallicity}
\label{abundance}

Both PoWR and BOSZ include stellar atmosphere grids for different values of the metallicity. BOSZ offers in total fourteen such options, and three of them, $[\mathrm{M/H}]=-1$, $[\mathrm{M/H}]=-1.75$, and $[\mathrm{M/H}]=-2.25$, are compared in Fig.~\ref{fig:Metallicity_6000K}. Here, JWST/NIRSpec mock spectra for $T_\mathrm{eff}=6000$ K are used, normalised to $m_\mathrm{AB}=26$ and focused on the Ca-H and Ca-K lines. Three Balmer lines shortward of the calcium lines in the figure have not been marked.

As calcium is an $\alpha$-element, the indirect effect of the metallicity can be studied by comparing the impact of the metallicity compared to the impact of the $\alpha$-abundance. The total abundance of the $\alpha$ elements can be written as $[\mathrm{\alpha/H}]=[\mathrm{\alpha/M}]+[\mathrm{M/H}]$. This equality was investigated by comparing the Ca-lines for one spectrum where $[\mathrm{M/H}]$ was increased with a certain value, with one where $[\mathrm{\alpha/M}]$ was also increased with the same value. This investigation showed that for a fixed effective temperature, the strength of the Ca-lines almost solely depends on the calcium abundance. Hereafter, although the metallicity is the parameter varied, the differences in the spectra will be credited to the Ca-abundance instead. 

In the Milky Way, low-metallicity stars are generally considered to be $\alpha$-enhanced, with [$\alpha$/M]$\sim0.4$ \citep[Fig. 4 in][]{Hayden15}. Generally, the scatter in abundances for metal-poor stars in the Milky Way is larger than for more metal-rich stars, with alpha-enhancements systematically higher than 0 \citep[Fig. 8 in][]{Lagae23}. Adopting [$\alpha$/M]$=0$ in this study is therefore a conservative approach, and the calcium lines might in actuality be stronger than seen here.

The Ca-abundance affects Ca-K at $\lambda=3934$ $\AA$ more than Ca-H at $\lambda=3969$ $\AA$, since the latter is blended with a hydrogen line. The strength of Ca-K decreases significantly with the Ca-abundance, making it easy to distinguish between spectra for $[\mathrm{M/H}]=-1$ and $[\mathrm{M/H}]=-1.75$. The difference between the $[\mathrm{M/H}]=-1.75$ and $[\mathrm{M/H}]=-2.25$ spectra is much smaller, and might be difficult to discern. In this case, it appears that a change of $\Delta[\mathrm{M/H}]=0.75$ is needed for the difference to be identifiable.

The Ca-K line might be detectable even at a very low metallicity, $[\mathrm{M/H}]=-2.25$, although the significance of this detection may be marginal (around $4\sigma$ at $z=3$). The significances with which these calcium lines are detected increase with redshift, and at $z=10$ and $[\mathrm{M/H}]=-2.25$ Ca-K is detectable with a significance of $\sim6\sigma$. For extremely metal-poor stars, with metallicities much lower than those included here, it is likely that no metal lines will be detectable.

Due to the fact that changing the metallicity is more or less equivalent to changing the calcium abundance (more specifically the $\alpha$-abundance) and that a change in metallicity causes a noticeable variation in the Ca-lines, Fig.~\ref{fig:Metallicity_6000K} suggests that the Ca-abundance of lensed stars can be roughly estimated using NIRSpec spectra. The results from the figure also implies that spectral lines in stars with quite low Ca-abundances might still be detectable, at least at the higher redshifts. To further constrain this abundance, the ratio of the two calcium lines can be used, as the Ca-H line is less affected than the Ca-K line. Since Ca-K is only detectable in a star with $T_\mathrm{eff}=6000$ K, and not at either 4000 K or 8000 K, a detection of the line can also help determine an approximate value of the effective temperature of the star. Unfortunately, lensed stars at 6000 K are likely yellow supergiants, a type of star with very few known examples. It therefore remains unclear whether such stars are likely to end up in a sample of lensed stars.

For the PoWR grids, only three values of the metallicity are available, equal to that of the Sun, the Large Magellanic Cloud (LMC), and the SMC. Generally, an increase in metallicity also increases the strength of the metal lines for these models, but a few particular absorption lines instead get weaker. The most interesting effect caused by increasing the metallicity is that emission lines appear. This only happens for stellar atmosphere models with $T_\mathrm{eff}=30\,000$ K and $[\mathrm{M/H}]=0$ (solar metallicity). Two emission lines in particular become detectable, the Bowen fluorescence lines \ion{O}{III} $\lambda\lambda$ 3133, 3444$\AA$ where the $\lambda3444$-line is only detectable at 26 AB mag with JWST, while the $\lambda3133$ $\AA$-line is also detectable at 27 AB mag. These lines are much weaker for $T_\mathrm{eff}=20\,000$ K, or with an LMC metallicity. Oxygen is, similar to calcium, an $\alpha$-element, meaning the metallicity effect on the spectral lines likely is due to the O-abundance. Since emission lines generally are easier to detect than absorption lines, they grant good prospects for observations. Also, considering that the lines are only detectable at a high metallicity, a detection of them in a lensed star would give a clear indication of its O-abundance.

One way to increase the significance of the metallicity measurements is by combining the information from several metal lines in the same spectrum, although this has not been attempted in the present study.

\subsection{Surface gravity limitations}
\label{logg}

For all stellar atmosphere models, the lowest available values of $\log(g)$ in the grids were chosen. Evolved stars with low surface gravity are expected to be favourably detected in samples of lensed stars \citep{Zackrisson23}, and many supergiants have been picked up in observations. However, not all evolved stars are covered by the PoWR-grid, in particular not those with Zero Age Main Sequence (ZAMS) masses $\gtrsim30-40 M_\odot$. To explore these supergiants, a grid reaching lower values of $\log(g)$ would be necessary. An example of such a grid, based on FASTWIND stellar atmosphere models, is described in \cite{Bestenlehner24}, where only spectral lines in the optical and NIR are included, and only for stars with $T_\mathrm{eff}\geq17\,000$ K. 

A slight increase in $\log(g)$, on the other hand, was implemented and did not provide any significant changes in the spectra. No lensed star with an extremely high value of $\log(g)$ was investigated in more detail, as such cases are deemed unlikely to be observed.

\subsection{Interstellar lines}

Certain detectable lines mentioned here can also be observed in absorption in the interstellar medium (ISM) of galaxies. In particular, the calcium H- and K-lines are usually strong in ISM clouds, and can be used to characterise the ISM if observed in supernova spectra \citep{Jenkins84}. Observations of strong Ca-H and Ca-K lines in spectra of lensed supernovae have been used to for example determine the redshifts of both the host- and lens-galaxies \citep{Goobar23, Johansson21}. In spectroscopic observations of lensed stars, there is a substantial risk that these ISM absorption lines from the host galaxy affect the spectra. Still, the stellar spectra are only affected if ISM is located in between the star and the observer. Some lensed stars may not be embedded in the host galaxy in such a way, and are thus not affected by the absorption in the ISM. An attempt to compensate for this effect can be done by spectroscopically studying other parts of the lensed arc.

\section{Conclusions}
\label{sec:conclusions}
Our conclusions can be summarised as follows
   \begin{enumerate}
            \item Multiple spectral lines will be detectable with both JWST/NIRSpec and ELT/HARMONI spectroscopy of lensed stars, assuming the star is observed with a peak magnitude of 26 AB mag. A few lines are also detectable at fainter magnitudes, up to 28 AB mag. At least for JWST, extremely long exposure times are necessary for these observations (50 hours per grating/filter pair), no matter which of the peak magnitude values is assumed. The number of observable lines changes with effective temperature and redshift, and the largest number of metal lines, as well as the strongest ones, appear at high effective temperatures ($T_\mathrm{eff}\geq15\,000$ K) and $z>3$. The strongest of these lines is \ion{C}{IV} $\lambda1550$ $\AA$ at $T_\mathrm{eff}=30\,000$ K. In addition, two strong calcium lines appear for $T_\mathrm{eff}=6000$ K and $z\geq3$. See Figs.~\ref{fig:Spectra6000-30000K} and \ref{fig:ELT_figures} as well as Tab~\ref{tab:MetalLines}.
            
            The observed lines might be useful to place constraints on for example the mass loss rate, terminal wind velocity, and effective temperature of the lensed stars (Sect.~\ref{Sec:use-of-features}). 
            \item ELT/HARMONI will be able to provide better spectroscopy, but at a more limited wavelength range, compared to JWST/NIRSpec. The strongest lines in the NIRSpec mock spectra may be detectable at magnitudes as faint as $\approx 28$ AB mag with HARMONI, even with much shorter exposure times, 5-10 hours per grating (Fig.~\ref{fig:ELT_figures}).
            \item No spectral lines are detectable for the coolest stars studied, with $T_\mathrm{eff}=4000$ K, as can be seen in Fig.~\ref{fig:Spectra4000-50000}. Although such stars are unlikely to appear in a sample of lensed stars at $z\geq6$, there is a bias towards observations of them at lower redshifts. A significant number of these stars might be possible to observe, but their spectra will not reveal much information.
			\item The mass loss rate of a lensed star has a large impact on the detectability of certain lines, at least for hot stars ($T_\mathrm{eff}\geq15\,000$ K) (Fig.~\ref{fig:Wind_30000K}). The emission component of the P Cygni profile of the \ion{C}{IV} $\lambda1550$$\AA$ line grows significantly in strength when the mass loss rate increases. If this component is observable for very faint stars (28 AB mag), the conclusion that the star has a high mass loss rate can be drawn.
            \item The metallicity also affects the strength of the spectral lines. This implies that certain metal abundances can be discerned with some accuracy with spectroscopic observations of lensed stars. The Ca-H and Ca-K lines for $T_\mathrm{eff}=6000$ K stars can be used as metal, particularly calcium, indicators and to help measure the effective temperature. Unfortunately, the 6000 K stars studied here are yellow supergiants, which is a very rare type of star. 
            \item For hot stars, $T_\mathrm{eff}=30\,000$ K, \ion{O}{III} Bowen fluorescence lines in rest-frame UV appear if the metallicity is sufficiently high. They only show up at solar metallicity, but are too weak to be detectable for a metallicity equal to that of the LMC. Emission lines are easier to detect than absorption lines, making their presence useful. Also, since they only appear at a high metallicity, their existence in an observed spectrum of a lensed star implies that the star has a high O-abundance.
            \item The faintest stars studied here, at 28 AB mag, have a few detectable emission lines, but no detectable absorption lines. The detectable emission lines include the \ion{C}{IV} $\lambda1550$ $\AA$-line for strong winds with JWST, or for the ELT with moderate mass loss rates. Another example is the \ion{He}{II} $\lambda1640$ $\AA$-line appearing in JWST-spectra of Wolf-Rayet stars.

   \end{enumerate}

In conclusion, it is possible to infer useful information about lensed stars using spectroscopic observations with JWST and ELT.

\begin{acknowledgements}
EZ acknowledges grant 2022-03804 from the Swedish Research Council, and has benefited from a sabbatical at the Swedish Collegium for Advanced Study.
AMA acknowledges support from the Swedish Research Council (VR 2020-03940). We thank the anonymous referee for their constructive report that helped improve the quality of this manuscript.
\end{acknowledgements}

\bibliographystyle{aa}
\bibliography{Lundqvist}

\newpage

\begin{table*}[bp] 
\renewcommand\thetable{A.1}
\centering
    \begin{threeparttable}
    \caption{Hydrogen and helium absorption lines detectable in mock JWST spectra at a peak magnitude of 26 AB mag.}
    \label{tab:HLines}
        \begin{tabular}{c|c|llll}
        \hline \hline 
        \noalign{\smallskip}
                         $T_\mathrm{eff}$ [K]    & $\log(g)$ [cgs]   & $z=1$ & $z=3$ & $z=6$                                                                                                                                                  & $z=10$  \\ \noalign{\smallskip} 
                                \Xhline{0.8pt}
        4000  & 0.0 & No detectable lines     &   No detectable lines    &    No detectable lines &  No detectable lines      \\ \hline
        
        6000  & 0.0 & \begin{tabular}[c]{@{}l@{}}\ion{H}{$\alpha$} $\lambda$6563$\AA$\end{tabular}     &   \begin{tabular}[c]{@{}l@{}}\ion{H}{$\eta$} $\lambda$3835$\AA$\\ \ion{H}{$\zeta$} $\lambda$3889$\AA$\\ \ion{H}{$\varepsilon$}  $\lambda$3970$\AA$\\ \ion{H}{$\delta$} $\lambda$4104$\AA$\\ \ion{H}{$\gamma$} $\lambda$4340$\AA$\\ \ion{H}{$\alpha$} $\lambda$6563$\AA$\end{tabular}    &    \begin{tabular}[c]{@{}l@{}}\ion{H}{I} $\lambda$3798$\AA$ \\ \ion{H}{$\eta$} $\lambda$3835$\AA$\\ \ion{H}{$\zeta$} $\lambda$3889$\AA$\\ \ion{H}{$\varepsilon$}  $\lambda$3970$\AA$\end{tabular}                                                               &  \begin{tabular}[c]{@{}l@{}}\ion{H}{I} $\lambda$3798$\AA$ \\\ion{H}{$\eta$} $\lambda$3835$\AA$\\ \ion{H}{$\zeta$} $\lambda$3889$\AA$\\ \ion{H}{$\varepsilon$}  $\lambda$3970$\AA$\\ \ion{H}{$\delta$} $\lambda$4104$\AA$\\ \ion{H}{$\gamma$} $\lambda$4340$\AA$ \end{tabular} \\ \hline
        8000  & 1.0 &  \begin{tabular}[c]{@{}l@{}}\ion{H}{$\beta$} $\lambda$4861$\AA$ \\ \ion{H}{$\alpha$} $\lambda$6563$\AA$\end{tabular}   &  \begin{tabular}[c]{@{}l@{}}\ion{H}{$\eta$} $\lambda$3835$\AA$\\ \ion{H}{$\zeta$} $\lambda$3889$\AA$\\ \ion{H}{$\varepsilon$}  $\lambda$3970$\AA$\\ \ion{H}{$\delta$} $\lambda$4104$\AA$\\ \ion{H}{$\gamma$} $\lambda$4340$\AA$\\ \ion{H}{$\beta$} $\lambda$4861$\AA$\\ \ion{H}{$\alpha$} $\lambda$6563$\AA$\end{tabular}     &  \begin{tabular}[c]{@{}l@{}}\ion{H}{I} $\lambda$3771$\AA$\\ \ion{H}{I} $\lambda$3798$\AA$\\ \ion{H}{$\eta$} $\lambda$3835$\AA$\\ \ion{H}{$\zeta$} $\lambda$3889$\AA$\\ \ion{H}{$\varepsilon$}  $\lambda$3970$\AA$\\ \ion{H}{$\delta$} $\lambda$4104$\AA$\\ \ion{H}{$\gamma$} $\lambda$4340$\AA$\\ \ion{H}{$\beta$} $\lambda$4861$\AA$\end{tabular}                                                                                                                                                      &    \begin{tabular}[c]{@{}l@{}}\ion{H}{I} $\lambda$3771$\AA$\\ \ion{H}{I} $\lambda$3798$\AA$\\ \ion{H}{$\eta$} $\lambda$3835$\AA$\\ \ion{H}{$\zeta$} $\lambda$3889$\AA$\\ \ion{H}{$\varepsilon$}  $\lambda$3970$\AA$\\ \ion{H}{$\delta$} $\lambda$4104$\AA$\\ \ion{H}{$\gamma$} $\lambda$4340$\AA$\end{tabular}     \\ \hline
        
        10\,000 & 2.0 & \begin{tabular}[c]{@{}l@{}}\ion{H}{$\alpha$} $\lambda$6563$\AA$\end{tabular}    &  \begin{tabular}[c]{@{}l@{}}\ion{H}{$\eta$} $\lambda$3835$\AA$\\ \ion{H}{$\zeta$} $\lambda$3889$\AA$\\ \ion{H}{$\varepsilon$}  $\lambda$3970$\AA$\\ \ion{H}{$\delta$} $\lambda$4104$\AA$\\ \ion{H}{$\gamma$} $\lambda$4340$\AA$\\ \ion{H}{$\beta$} $\lambda$4861$\AA$\\ \ion{H}{$\alpha$} $\lambda$6563$\AA$\end{tabular}     &\begin{tabular}[c]{@{}l@{}}\ion{H}{I} $\lambda$3771$\AA$\\ \ion{H}{I} $\lambda$3798$\AA$\\ \ion{H}{$\eta$} $\lambda$3835$\AA$\\ \ion{H}{$\zeta$} $\lambda$3889$\AA$\\ \ion{H}{$\varepsilon$}  $\lambda$3970$\AA$\\ \ion{H}{$\delta$} $\lambda$4104$\AA$\\ \ion{H}{$\gamma$} $\lambda$4340$\AA$\\ \ion{H}{$\beta$} $\lambda$4861$\AA$\end{tabular}                                                                                                                                                      &    \begin{tabular}[c]{@{}l@{}}\ion{H}{I} $\lambda$3771$\AA$\\ \ion{H}{I} $\lambda$3798$\AA$\\ \ion{H}{$\eta$} $\lambda$3835$\AA$\\ \ion{H}{$\zeta$} $\lambda$3889$\AA$\\ \ion{H}{$\varepsilon$}  $\lambda$3970$\AA$\\ \ion{H}{$\delta$} $\lambda$4104$\AA$\\ \ion{H}{$\gamma$} $\lambda$4340$\AA$\end{tabular}     \\ \hline
        15\,000 & 2.0 & No detectable lines     &   No detectable lines    &   No detectable lines                                                                                                                                                     &   No detectable lines     \\ \hline
        
        20\,000 & 2.4 & No detectable lines      & No detectable lines       & No detectable lines &   No detectable lines      \\ \hline
        
        30\,000  & 3.2 &No detectable lines      & No detectable lines      & No detectable lines                                                                                                                                                        &     No detectable lines   \\ \hline
        
         50\,000  & 4.2 & No detectable lines     &   No detectable lines    &    No detectable lines &  No detectable lines     \\
        \hline
         \begin{tabular}[c]{@{}c@{}} 50\,000\\ (WR-star) \end{tabular} & 4.0 & \begin{tabular}[c]{@{}l@{}}\ion{He}{II} $\lambda$5411$\AA$\\ \ion{He}{II} $\lambda$6560$\AA$\\ \ion{He}{II} $\lambda$10130$\AA$ \\ \ion{He}{II} $\lambda$11630$\AA$ \\ \ion{He}{II} $\lambda$18636$\AA$  \end{tabular}  &   
         
        \begin{tabular}[c]{@{}l@{}}\ion{He}{II} $\lambda$2733$\AA$\\ \ion{He}{II} $\lambda$3203$\AA$\\ \ion{He}{II} $\lambda$4686$\AA$ \\ \ion{He}{II} $\lambda$6560$\AA$ \\ \ion{He}{II} $\lambda$10130$\AA$  \end{tabular}   &    
         
        \begin{tabular}[c]{@{}l@{}}\ion{He}{II} $\lambda$1640$\AA$ \\ \ion{He}{II} $\lambda$2733$\AA$ \\ \ion{He}{II} $\lambda$3203$\AA$ \\ \ion{He}{II} $\lambda$4686$\AA$\end{tabular} &  
         
         \begin{tabular}[c]{@{}l@{}}\ion{He}{II} $\lambda$1640$\AA$ \\ \ion{He}{II} $\lambda$3203$\AA$\end{tabular}     \\
        \hline
        \multicolumn{1}{c}{}\vspace{5cm}
        
        \end{tabular}
    \end{threeparttable}
\end{table*}
~
\newpage

\section*{Appendix A: Detectable hydrogen and helium lines with JWST}

The hydrogen and helium absorption lines detectable with JWST/NIRSpec for the used combinations of effective temperatures and redshifts are presented in Table~\ref{tab:HLines}. The lines are detectable at 26 AB magnitudes, but none of the hydrogen lines are detectable at 27 AB mag, with an exposure time of 50 hours per grating/filter. All included hydrogen lines are Balmer lines, and only for the lower effective temperatures, $T_\mathrm{eff}\leq15\,000$ K, are they strong enough to be detectable. For the hotter stars, the lines might be present in a noise-free spectrum, but not discernible when noise has been included. The 50\,000 K WR-star is, additionally, hydrogen free, meaning that no hydrogen lines are present. Instead, multiple strong helium lines are detectable, as expected for a WR-star. Many of these lines are also detectable at 27 AB mag.

\end{document}